\documentclass[isre,nonblindrev]{informs3} 

\DoubleSpacedXI

\usepackage{natbib}
 \bibpunct[, ]{(}{)}{,}{a}{}{,}%
 \def\bibfont{\small}%

\usepackage{graphicx}

\usepackage{tabularx}  
\usepackage{tabulary}
\usepackage{multirow}
 \usepackage{booktabs}

\usepackage{xcolor}
\usepackage{amsmath}

\usepackage{subcaption}
\usepackage{graphicx}
\TheoremsNumberedThrough     
\ECRepeatTheorems

\EquationsNumberedThrough    

\MANUSCRIPTNO{}

\begin{document}



\RUNTITLE{}

\TITLE{Customizing Large Language Models for Business Context: Framework and Experiments}

\ARTICLEAUTHORS{\AUTHOR{
 Wen Wang$^1$ \quad Zhenyue Zhao$^2$\quad Tianshu Sun$^3$\\
  $^1$University of Maryland\quad$^2$Cornell University\quad$^3$University of Southern California  \\
  \texttt{wenw@umd.edu}, \texttt{njujack@163.com}, \texttt{tianshus@usc.edu} }}


\ABSTRACT{%



The advent of Large Language Models (LLMs) has ushered in a new era for design science in Information Systems, demanding a paradigm shift in tailoring LLMs design for business contexts. We propose and test a novel framework to customize LLMs for general business contexts that aims to achieve three fundamental objectives simultaneously: (1) aligning conversational patterns, (2) integrating in-depth domain knowledge, and (3) embodying theory-driven soft skills and core principles. We design methodologies that combine domain-specific theory with Supervised Fine Tuning (SFT) to achieve these objectives simultaneously. We instantiate our proposed framework in the context of medical consultation. Specifically, we carefully construct a large volume of real doctors’ consultation records from a leading online medical consultation platform and a vast amount of medical knowledge from multiple professional databases. Additionally, drawing on medical theory, we identify three soft skills and core principles of human doctors: professionalism, explainability, and emotional support, and design approaches to integrate these traits into LLMs. We demonstrate the feasibility of our framework using online experiments with thousands of real patients as well as evaluation by domain experts and consumers. Experimental results show that the customized LLM model substantially outperforms untuned base model in medical expertise as well as consumer satisfaction and trustworthiness. It significantly narrows the gap between untuned language models and human professionals, significantly moving closer to human-level performance. Additionally, we delve into the characteristics of textual consultation records and adopt interpretable machine learning techniques to identify what drives the performance gain. Finally, we showcase the practical value of our model through a decision support system designed to assist human doctors in a lab experiment with medical professionals. We deploy an online consultation platform and use our model to generate initial responses, aiding doctors in a copilot fashion during consultations. We find that the customized LLM model enhances doctors’ productivity by 53.16\% in terms of time spent and reduces the cognitive load for human doctors during consultations, making it highly effective and suitable for large-scale implementation in practice. 
}%



\maketitle

%

\newpage
\section{Introduction}

In the ever-evolving landscape of Information Systems, design science has played a pivotal role in shaping how to tailor technical designs to address complex business needs. This field has witnessed a wide range of customization and application of various Artificial Intelligence (AI) models for business contexts and needs. These models range from recommender systems~\citep{zhou2023longitudinal, zhang2020consumption, yin2022diversity} to natural language processing (NLP)~\citep{lee2018advertising, liu2019go, manzoor2023influence, xie2022understanding}, forecasting and predictions~\citep{li2023predicting, sun2022predicting, macha2023personalized, ben2020trajectories, liu2020predicting}, and optimization algorithms~\citep{wang2023deep, song2023ensemble, kokkodis2021demand}.


The recent advancement of Large Language Models (LLMs) marks a significant milestone in AI, showing transformative improvement over traditional AI models.  LLMs revolutionize and significantly differ from traditional AI models in four key ways. Firstly, they are \textit{general-purpose} AI models, adept at a wide variety of tasks such as language understanding, generation, and even reasoning. This contrasts with traditional AI systems, which are generally task-specific, with models optimized for singular tasks.  Second, LLMs have superior scale and learning capabilities, capable of absorbing vast industry data at the trillion level. Third, LLMs specialize in understanding unstructured data (i.e., text), whereas traditional AI systems tend to  gear towards processing structured data.  Finally, LLMs revolutionize user interaction by enabling end-to-end conversational engagements, closely simulating real-world human conversations. This represents a significant advance over the rigid, preset rules, or classifications-based communication of previous AI systems.

Such transformative advancement presents a substantial challenge to the traditional research paradigm in design science. The conventional approach, characterized by defining specific tasks or goals like predicting ratings or informativeness levels and creating datasets and models to link inputs and outputs for these tasks, is becoming increasingly inadequate  in the face of these new developments. The emerging capabilities of LLMs enable the development of highly customized solutions to function as  professional experts with extensive domain expertise. These experts can handle a broad range of tasks within their domains, offering a new generation of user experience services through human-like  conversational interactions. Given these new capabilities and the expanded scope they offer, there is an urgent need for a paradigm shift towards redefining and adjusting technical designs to better align with business contexts in the LLM era.

Addressing this gap, our study introduces a novel framework designed to tailor technical design to business contexts in the LLM era.  We are among the first to demonstrate how to customize LLMs for general business contexts in Information Systems.  Leveraging LLMs' superior capabilities, it enables customization for various business contexts, such as customer support, medical consultation, legal assistance, sales and marketing, educational programs, and more. Our goal is to provide a versatile framework applicable across any business context, marking a significant step forward in aligning technical advancements with practical business needs.


Our proposed framework consists of two steps. Firstly, we decompose the business value of customizing LLM for a business context into three fundamental objectives: $(i)$ alignment with conversational patterns, $(ii)$ in-depth domain knowledge, $(iii)$ soft skills and core principles of a professional role. These objectives represent the three major gaps between LLMs and a specific business context and are the main areas we need to customize. Secondly, we design methodologies to model these objectives simultaneously by combining domain-specific theory with Supervised Fine Tuning (SFT) in large language models.

\begin{figure}
\centering
\begin{subfigure}[b]{0.495\textwidth}
\centering
\includegraphics[width=\textwidth]{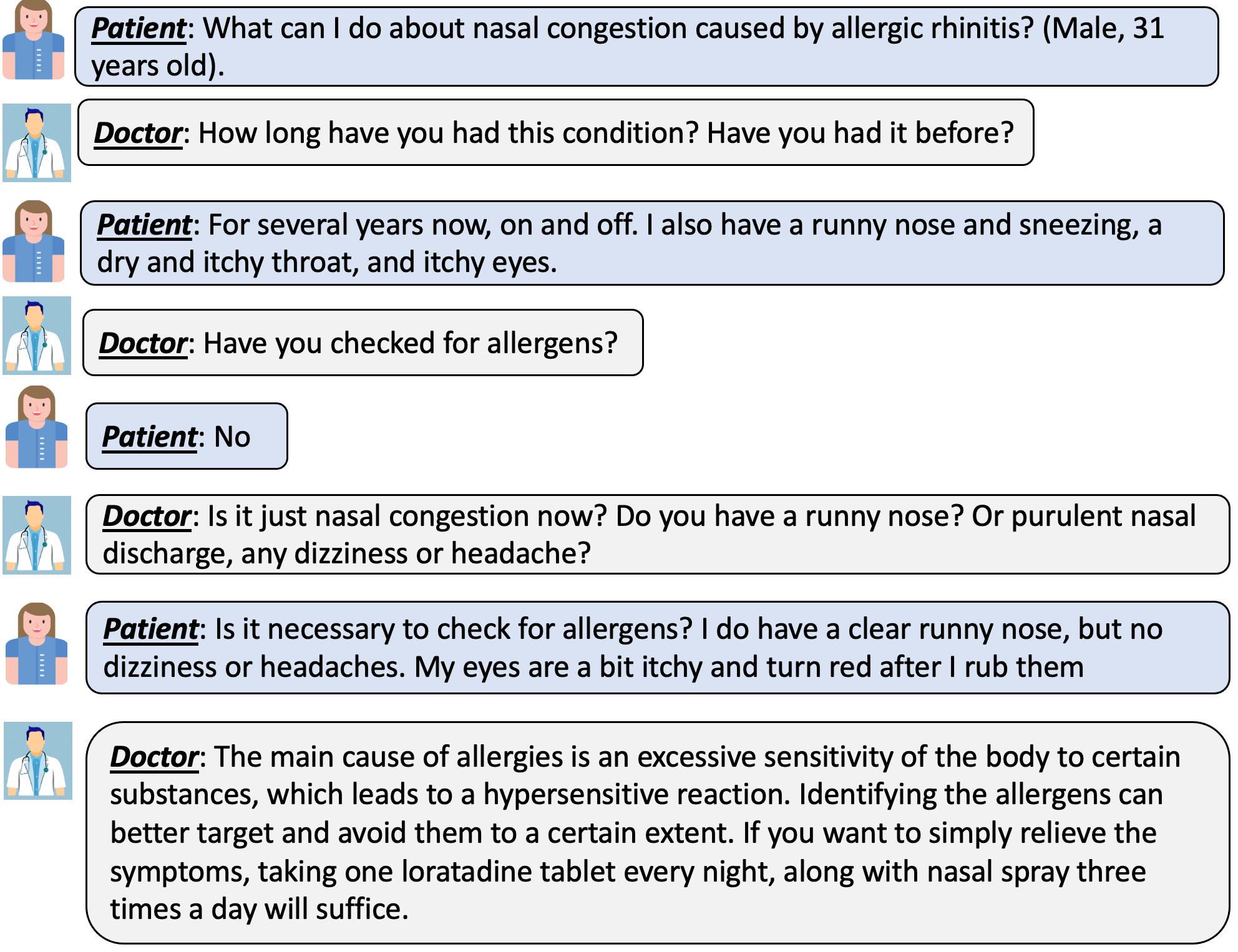}
\caption{ Real Doctor Record}
\label{fig:record_example}
\end{subfigure}
     \hfill
\begin{subfigure}[b]{0.495\textwidth}
\centering
\includegraphics[width=\textwidth]{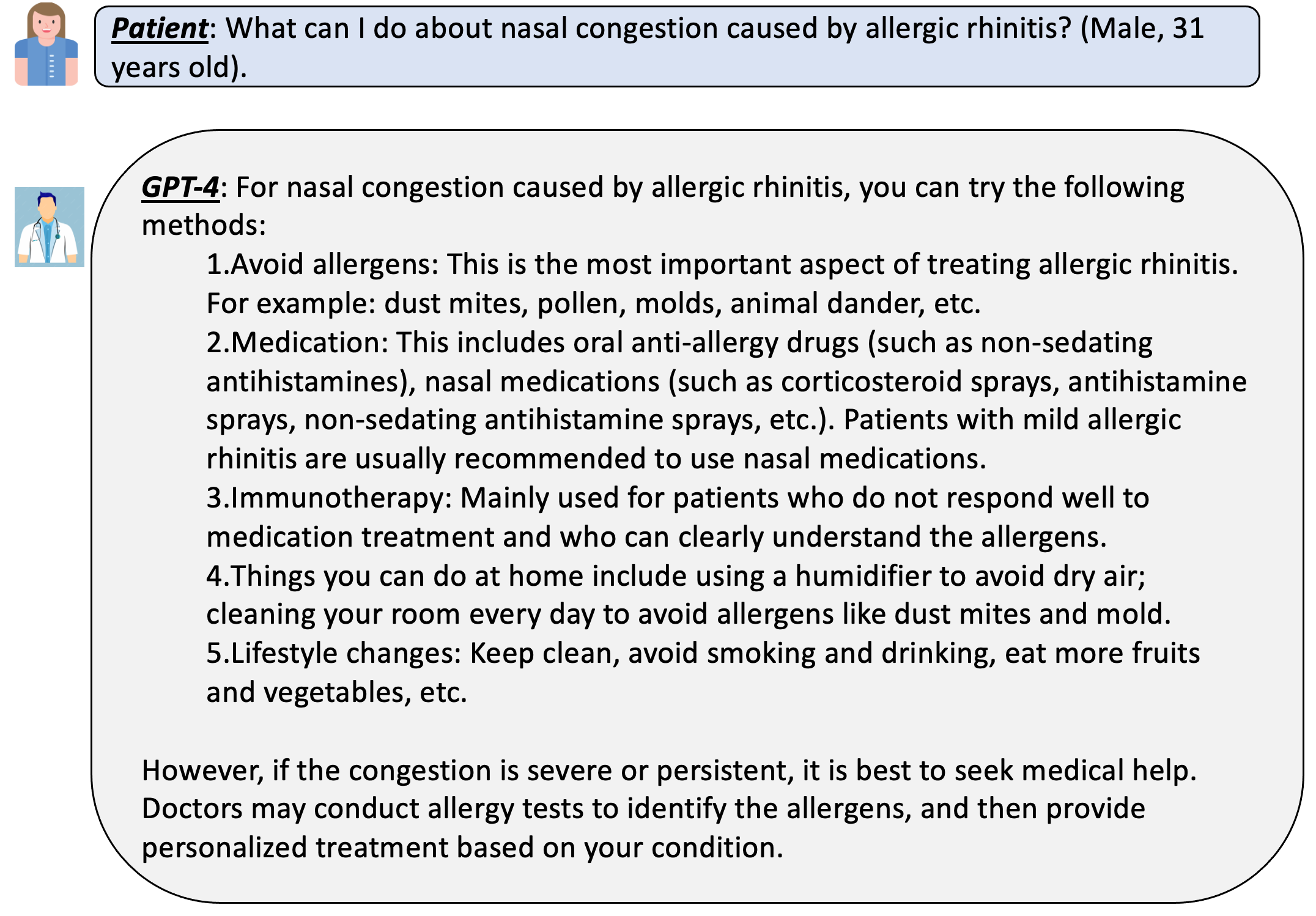}
\caption{GPT-4 Response to Patient's Inquiry}
\label{fig:record_example_gpt}
\end{subfigure}
\caption{Real doctor vs. GPT-4 Response to Patients' Inquiry}
\label{fig:academic_attention}
\end{figure}

The decomposed three fundamental objectives as follows: 

$(i)$ \textit{Alignment with conversational patterns}: Existing LLMs, such as GPT-4, deviate from the conversational patterns typical of a professional role in a specific business context. Each business domain possesses its unique conversation pattern with specific communication  structures and  norms,  reflecting its deep business needs and domain-specific techniques. Taking medical consultation as an example (See Figure~\ref{fig:record_example} for an example of consultation record), human doctors typically conduct multiple conversational turns to first gather adequate patient information and deeply understand the health conditions, then offer diagnoses and treatment recommendations, with each conversation having a distinct intention.  In contrast, existing LLMs primarily act as single-turn agents with limited multi-turn questioning capabilities concerning domain-specific requirements (e.g., a user's health specifics in the context of medical consultation). For example, Figure~\ref{fig:record_example_gpt} shows a GPT-4 response to a patient inquiry, which largely deviates from the  human doctors' pattern and often includes many irrelevant details.  Failing to align with the conversational patterns of professional roles can compromise reliability and specificity, potentially undermining consumer trust and experience.




$(ii)$ \textit{In-depth domain knowledge}: 
The use of LLMs in specific business domains  must uphold an exceptionally high level  of domain-specific knowledge. This is crucial for all contexts to ensure the accuracy and reliability of the information and recommendations provided. Moreover, it is particularly critical in high-stake fields, such as healthcare, legal, and education, where insufficient domain knowledge can lead to serious consequences. In healthcare, for example, without thorough medical knowledge, AI systems could potentially offer misdiagnoses or suggest inappropriate treatments, severely  endangering patient safety.  Similarly, in the legal domain, insufficient legal knowledge could lead to incorrect conclusions or advice, potentially resulting in serious legal ramifications. Thus, maintaining precise and deep domain knowledge is crucial for the effectiveness and trustworthiness of LLMs in these specialized contexts.

$(iii)$ \textit{Soft skills and core principles of professional roles}: Beyond domain knowledge and conversational pattern alignment, it is crucial for LLM to emulate the soft skills and core principles characteristic of various business domains. This includes non-technical skills, interpersonal competencies, and foundational principles essential for professional excellence. This aspect represents distinct competencies separate from the previous two objectives. Soft skills and core principles focus on human-centric abilities, while conversational pattern alignment involves adhering to specific communication structures and norms within a particular professional context and in-depth domain knowledge pertains to the technical or specialized knowledge required in a field. For example, in the medical field, theory suggests that skills such as professionalism, clear explanations, and emotional support are essential for patient satisfaction and building trust with patients~\citep{wilkinson2009blueprint,ha2010doctor,markides2011importance,walter2021partnership}. These soft traits contribute significantly to a positive consumer experience and are critical for LLMs to offer an engaging, satisfactory, and trustworthy experience in each business domain.

To simultaneously achieve three objectives, we design methodologies to integrate domain-specific
theory with Supervised Fine Tuning  in LLMs.  Our approach involves creating a comprehensive SFT dataset, divided into three distinct parts to meet these objectives: $(i)$ A substantial collection of real-world professional conversation records. This aids LLMs in learning and mimicking the conversational patterns of actual professionals; $(ii)$ Utilization of domain-specific theories to pinpoint essential soft skills and core principles professionals should embody. We leverage this to select exemplary conversation records, enabling LLMs to demonstrate these skills and principles; $(iii)$  A vast array of domain knowledge in the form of question and answer pairs, enhancing the LLMs' ability to maintain a high level of domain-specific knowledge.
Following dataset construction, we conduct supervised fine-tuning on open-source LLMs using this dataset. Our method employs the LoRA (Low-rank Adaptation) fine-tuning strategy, which is widely recognized for its effectiveness~\citep{hu2021lora}. 


To validate the proposed framework, we instantiate our framework in the context of medical consultation. 
The goal is to customize LLMs to closely resemble professional doctors, known as LLM-doctors, thereby enabling them to potentially assist human doctors in practice.  AI-assisted medical consultations hold tremendous value for the healthcare industry, influencing both patient experiences and the efficiency of healthcare systems. This industry has experienced rapid growth and increasing demand in recent years.  According to \citet{medicalbotsize2_2023} and \citet{medicalbotsize3_2023}, the market was valued at USD 211.60 million in 2022 and is projected to reach USD 647.29 million by 2030, growing at a compound annual growth rate (CAGR) of 15\% over the forecast period. 

Specifically, we carefully construct a multi-source professional knowledge database for LLMs customization in this context. We collect a large volume of real doctors' consultation records from a leading online medical consultation platform and a vast amount of medical knowledge from multiple professional databases.  Additionally, drawing on medical theory, we identify three key soft skills and core principles of human doctors: professionalism, explainability, and emotional support, and  design approaches to integrate these theory-driven skills  into LLMs. 

Then, we adopt multiple experiments, including online experiment and lab experiment, to demonstrate the effectiveness and practical value of our framework.  We conduct a large scale of online experiment with thousands of real patients to conduct medical consultations, which is followed by comprehensive evaluations regarding medical expertise by medical professionals as well as  consumer preferences by real consumers, demonstrating the effectiveness and feasibility of our framework. Additionally, we delve into the characteristics of textual consultation records and adopt interpretable machine learning techniques to identify what drives the performance gain. Finally, we showcase the practical value of our model through a decision support system designed to assist human doctors in a lab experiment with medical professionals.

Our results reveal several compelling findings:

Firstly, we conduct an online experiment with thousands of real patients who had recently felt unwell and sought medical attention to evaluate the effectiveness of our model. We compare the performance of LLM-doctor with untuned base model as well as human doctors. Our evaluations from medical expertise perspective, conducted by medical professionals, show that LLM-doctor significantly outperforms the untuned base model with an +11.68\% improvement in professionalism ($p < 0.001$) and a +17.43\% improvement in accuracy ($p < 0.001$). Meanwhile, the base model significantly lags behind human doctors by -13.08\% in professionalism ($p < 0.001$) and by -16.9\% in accuracy ($p < 0.001$). Our LLM-doctor more closely matches the medical expertise of human doctors, with marginally smaller gaps of -2.93\% in professionalism ($p = 0.004$) and -2.42\% in accuracy ($p = 0.019$). This indicates that our framework significantly narrows the gap between untuned LLMs and human doctors, bringing it closer to human-level performance. This result is particularly significant, as a single LLM-model can approach the capabilities of  thousands of human doctors across different outpatient departments. 

The same conclusion applies to consumer preference metrics evaluated by real consumers. From a consumer satisfaction perspective, the untuned base model performs -27.76\% worse than human doctors ($p < 0.001$), and from a consumer trustworthiness perspective, it is -27.16\% worse ($p < 0.001$). The gap in consumer preferences between the base and human doctors is much larger than that in medical expertise. This indicates that the untuned base model significantly lags behind when engaging with consumers in reality, highlighting a potential shortcoming. This also underscores the need for a framework to improve in this aspect. From this standpoint, LLM-doctor significantly outperforms the base model with a +28.36\% improvement in consumer satisfaction ($p < 0.001$) and a +27.56\% improvement in trustworthiness ($p < 0.001$). It narrows the gap to human doctors' medical expertise with smaller deficits of -7.27\% in satisfaction ($p < 0.001$) and -7.09\% in trustworthiness ($p < 0.001$). This further indicates that our framework significantly narrows the gap between untuned LLMs and human doctors, moving it closer to human-level performance, with an even stronger improvement from the perspective of consumer preference.

Furthermore, beyond model comparison, we conduct an in-depth interpretability analysis using interpretable machine learning techniques to understand what happens in conversation records and where the performance gains come from. We specifically examine which conversational or communication characteristics contribute to these gains. We extract various textual communication and consultation characteristics, such as whether the consultation  demonstrates proactive information gathering, provides targeted disease diagnosis, offers targeted treatment recommendations, uses medical terms, engages in multi-round interactions, exhibits a human-like communication style, demonstrates clarity of response, and shows respect and patience in communication, as well as logical reasoning. We find that the LLM-doctor is much closer to human doctors across the majority of these textual communication and consultation characteristics, whereas the base model significantly lags behind. This indicates that our framework can profoundly change how LLMs conduct consultations, including the procedures and communication styles, making them closer to those of human doctors.

Subsequently, we apply interpretable machine learning model to use the extracted textual characteristics to predict perceived professionalism, accuracy, consumer trustworthiness, and satisfaction. Interestingly, we find that the characteristics contributing to perceived medical expertise and consumer preference are quite distinct. For example, in terms of accuracy, the top predictors are targeted disease diagnosis, medical term usage, proactive information gathering, multi-round interaction, and targeted treatment recommendation. For consumer satisfaction, the top predictors are targeted treatment recommendations, respect and patience in communication, and clarity of response, among others. This analysis not only confirms the effectiveness of our model but also enhances our understanding of the essential elements of successful medical consultations from both medical expertise and consumer experience perspectives.

Last but not least, we demonstrate the practical value of our model by illustrating how it can assist human doctors in the real world. Specifically, we showcase a decision support use case in which, during an online consultation, our LLM-doctor can generate an initial response to patients' queries in a copilot manner. The human doctor can either accept this generated response or edit it before finalizing and sending it back to the patients. The copilot assistant approach enables human doctors to lead the entire process, maintain control over the final response and conduct quality assurance, thereby mitigating potential errors or risky advice generated by the AI model.  To achieve this goal, we  develop an online consultation platform where our LLM-doctor or base model can be embedded in the backend, and we conduct a laboratory experiment, inviting experienced doctors from prestigious hospitals in China to conduct medical consultations with this platform. Results show that with the aid of the LLM-doctor, human doctors, on average, saved 53.16\% of their time compared to operating without the LLM-doctor during consultation. The base model also saves human doctors' time. However, the magnitude is much smaller compared to the LLM-doctor; it saved only 19.31\% of the time compared to doctors working independently. This suggests that our LLM-doctor  has larger potential to enhance the productivity of human doctors compared to the base model.

Additionally, through a post-hoc survey for human doctors to evaluate their experience with the LLM-doctor or untuned base model, interestingly, the results show that human doctors generally perceive the response quality of the LLM-doctor to be higher, and they adopt its responses (either directly or by modifying, retaining at least 50\% of the original content) more frequently than those of the base model. Additionally, the doctors believe that the LLM-doctor can reduce their cognitive load during consultations much more than the base model. They also think that the LLM-doctor can save them time spent during consultations more effectively than the base model. Furthermore, they perceive the LLM-doctor as having more practical value for large-scale deployment on online consultation platforms compared to the base model.

Furthermore, through open-ended feedback from human doctors, overall, human doctors think the base model has salient drawbacks, including providing overly verbose and generic responses, a lack of personalization, and many irrelevant details which might increase anxiety for patients. These factors make it less useful for doctors and patients. In contrast, our LLM-doctor can proactively ask questions and address the key points in consultations and can be directly used by human doctors during consultations. All these factors demonstrate that our LLM-doctor’s practical value is greater than that of the base model.

Overall, our contributions are fourfold. Firstly, we propose a novel framework to customize LLMs for general business contexts, aiming to achieve three fundamental objectives simultaneously: (1) aligning conversational patterns, (2) integrating in-depth domain knowledge, and (3) embodying theory-driven soft skills and core principles. We design methodologies that combine domain-specific theory with Supervised Fine Tuning (SFT) to achieve these objectives. Our framework facilitates the customization of LLMs to serve as general-purpose professional experts for business purposes, enhancing domain expertise, consumer satisfaction, and trustworthiness. Secondly, we demonstrate the effectiveness of our proposed framework in the context of medical consultations with a large-scale online experiment. Our fine-tuned LLM-doctor model significantly outperforms the untuned base model and substantially reduces the gap between untuned LLMs and human doctors, elevating LLMs to the level of human experts. Thirdly, we conduct an interpretability analysis to identify what drives performance gains and the interpretable insights gained enhance our understanding of the essential elements of successful customization from both technical expertise and consumer experience perspectives. Finally, we showcase how to use our model in practice and demonstrate its practical value through a decision support system designed to assist human doctors in a lab experiment with medical professionals. In summary, our proposed framework offers step-by-step principles, guidance, and valuable insights for future research on customizing LLMs to address real-world business problems.

\section{Literature Review}
Our work builds upon the extensive literature on design science, the emerging topic of Generative AI in information system literature, as well as the emerging literature on customizing LLMs for real-world applications, all of which we briefly review in this section.

\subsection{Design Science in Information Systems }

Our study is closely related to design science studies  in Information System (IS) literature. The field of design science has been instrumental in influencing the way technology meets the demands of intricate business requirements. Within this realm, we have observed a diverse array of adaptations and implementations of different AI models to tackle diverse business queries. One notable area of focus in existing literature is the development of recommendation systems aimed at fulfilling specific business needs. For instance, \citet{yin2022diversity} address the limitations of traditional recommendation systems that tend to suggest similar options, by proposing a diversity preference-aware link recommendation model. Similarly, \citet{zhou2023spoiled} present a personalized healthcare recommendation framework, enhancing individualized wellness by matching users with suitable health interventions.

Another significant area of research involves customizing natural language processing models for various business applications. For example, \citet{lee2018advertising} develop an NLP model to evaluate the content of brand advertisements on social media platforms like Facebook, focusing on informativeness and brand personality traits. \citet{liu2019go} tailor a bidirectional long short-term memory model for identifying medical terms in YouTube's healthcare educational videos, categorizing these videos based on the extent of medical information they present. Furthermore, \citet{xie2022understanding} and \citet{yang2023getting} explore the realms of social media and psycholinguistics respectively, with the former developing a sentiment-enriched deep learning method to analyze medication nonadherence from social media posts, and the latter proposing an NLP-based approach for personality detection.

Additionally, a third strand of literature focuses on forecasting, prediction, and optimization in various contexts. This includes \citet{sun2022predicting} using deep learning to anticipate consumers' future purchasing paths by analyzing their omnichannel behaviors, \citet{macha2023personalized} introducing a personalized privacy preservation framework for consumer mobile location data, and \citet{ben2020trajectories} developing predictive models for early risk assessment in chronic disease patient readmissions. Moreover, \citet{chen2023theory} and \citet{wang2023deep} contribute with a theory-driven method to predict customer responses in a specific commercial mode and a deep reinforcement learning framework for optimizing sequential targeting, respectively.

The existing body of work typically adheres to paradigms that define specific tasks or goals, subsequently developing datasets and models to connect inputs with outputs for these tasks. However, with the transformative capabilities of LLMs, such paradigms are increasingly facing challenges.

Our study contributes to this field by introducing a new framework that aligns technical design with business contexts in the LLM era. We are among the first to decompose general business value into three fundamental objectives and design methodologies to achieve these objectives simultaneously, customizing LLMs for broad business applications in Information Systems. Our framework facilitates the customization of LLMs to serve
as general-purpose professional experts for business purposes, including demonstrating domain
expertise and enhancing consumer satisfaction and trustworthiness.  The customized LLMs possess extensive domain knowledge, along with the necessary soft skills and core principles, aligning their communication patterns with those of domain professionals. These experts are capable of handling a wide array of tasks within their domains through human-like conversational interactions. We validate the effectiveness and performance of our framework through its application in medical consultations. Additionally, we delve into the characteristics of textual consultation records and adopt interpretable machine learning techniques to identify what drives the performance gain. Finally, we showcase the practical value of our model through a decision support system designed to assist human doctors in a lab experiment with medical professionals. Our framework and findings offer step-by-step principles and practical insights for future research on customizing LLMs to solve real-world business needs.

\subsection{Generative AI}

Our study is also closely relevant  to the growing body of work on generative AI within IS.  The area of AI, particularly LLMs, has experienced remarkable advancements, heralding a new era in AI capabilities. LLMs like GPT-4, trained on a vast collection of data, demonstrate capabilities beyond advanced language processing; they display elements of broader intelligence. GPT-4, in particular, has shown exceptional performance, equating to human standards in various professional and academic tests, such as the Uniform Bar Exam, SAT, GRE, and AP free-response questions~\citep{GPT4_2023, bubeck2023sparks, zhang2023one, liu2023summary, yang2023harnessing}.

The significance of Generative AI has prompted a multitude of investigations into its implications and practical uses. \citet{eloundou2023gpts} suggests that GPTs might influence approximately 80\% of job tasks in the U.S. labor market. Research by \citet{noy2023experimental} and \citet{brynjolfsson2023generative} reveals that ChatGPT substantially improves average productivity, often substituting for human labor instead of complementing skills. According to \citet{zhou2023generative}, while generative AI boosts output in design fields, it might diminish creative abilities. Investigations by \citet{wang2023s} and \citet{wang2023unraveling} delve into the business prospects of AI-generated imagery, especially those emulating artistic styles, and offer a method for assessing the intelligence of LLMs. Additionally, \citet{burtch2023consequences} demonstrates how ChatGPT is replacing human contributions in online information communities, emphasizing the necessity of social interaction to mitigate the risks associated with AI.

Our research contributes to this area by focusing on  tailoring of LLMs for addressing business-related queries from a design science and technical standpoint. We introduce a new framework for customizing LLMs for general business contexts usages, demonstrating its effectiveness and feasibility. Our work not only presents a feasible model for applying LLMs in business contexts but also delves into the technical intricacies of model development, offering practical solutions for customizing LLMs for real-world business issues.

\subsection{Customize LLMs for Real-world Applications}

The rapid progress of LLMs  has sparked a wave of innovation, leading to the development of customized LLMs for a diverse range of specific applications. For instance, \citet{reisenbichler2022frontiers} have fine-tuned a GPT model specifically for the niche of content marketing, focusing on the creation of SEO-optimized landing pages.  Similarly, \citet{liga2023fine} have taken these models into the legal domain, adapting a GPT model for the intricate task of legal rule classification. In the educational sector, \citet{fan2023grammargpt} have developed LLMs specifically designed for writing and grammar assistance.  Finally, \citet{zhang2023benchmarking} have tailored LLMs for the task of news summarization.

Additionally, customizing LLMs to support health service is also emerging area. For example, \citet{li2023chatdoctor} introduce a medical assistant adapted from the LLaMA model, trained with real patient-doctor dialogue data. \citet{han2023medalpaca} develop MedAlpaca by integrating Stanford Alpaca and AlpacaLoRA technologies to enhance medical question-answering and dialogue capabilities. Further, \citet{wu2023pmc} use medical papers to refine medical assistant's performance in medical tasks. \citet{singhal2023large} introduce  Med-PaLM, demonstrating remarkable effectiveness in various benchmark tests. 




We contribute to this literature by presenting a novel and systematic framework to integrate comprehensive business value into LLMs, including both technical expertise  and non-technical  aspects augmentation. Oriented toward business value objectives, our model enhances domain expertise, consumer satisfaction, and trustworthiness. It not only enhances medical expertise but also improves its ability to engage consumers and build trust. This dual focus on technical proficiency and consumer experience represents a significant leap forward from existing approaches, which often emphasize one aspect over others and typically neglect the business value of non-technical aspects augmentation, thus overlooking consumer satisfaction and trustworthiness aspects. For instance, \citet{singhal2023large, wu2023pmc, han2023medalpaca} enhance medical knowledge but fall short in aligning with the conversation patterns, soft skills, and core principles of real doctors.  Similarly, \citet{li2023chatdoctor} and \citet{bao2023disc} utilize real doctors' conversation records, but they do not explicitly model and capture the soft skills and core principles of human doctors.  Additionally, the previous method is more domain-specific, whereas our framework is a general and systematic framework that can be applied to any business context. Our proposed framework offers step-by-step principles and guidance on customizing LLMs to address real-world business problems, enabling customization for various business contexts, such as customer support, medical consultation, legal assistance, sales and marketing, educational programs, and more.

\section{ Proposed Framework and Methodologies }

In this section, we introduce our proposed framework along with its methodologies, specifically applying them to the context of medical consultation. Our framework is structured into two main steps. The first step involves decomposing the overarching business value into three primary objectives. These objectives are crucial for ensuring that the framework is tailored to the unique needs and challenges of the given context. The second step focuses on the methodologies employed to achieve these objectives. Here, we combine domain-specific theories with Supervised Fine Tuning  techniques in large language models. 

In the specific case of medical consultation, our aim is to tailor LLMs to function similarly to professional doctors. This customized LLMs, referred to as ``LLM-doctor,'' is designed to assist general-purpose medical consultation through human-like conversational interactions.

\begin{figure}[tbh!]
\centering
\includegraphics[scale = 0.35]{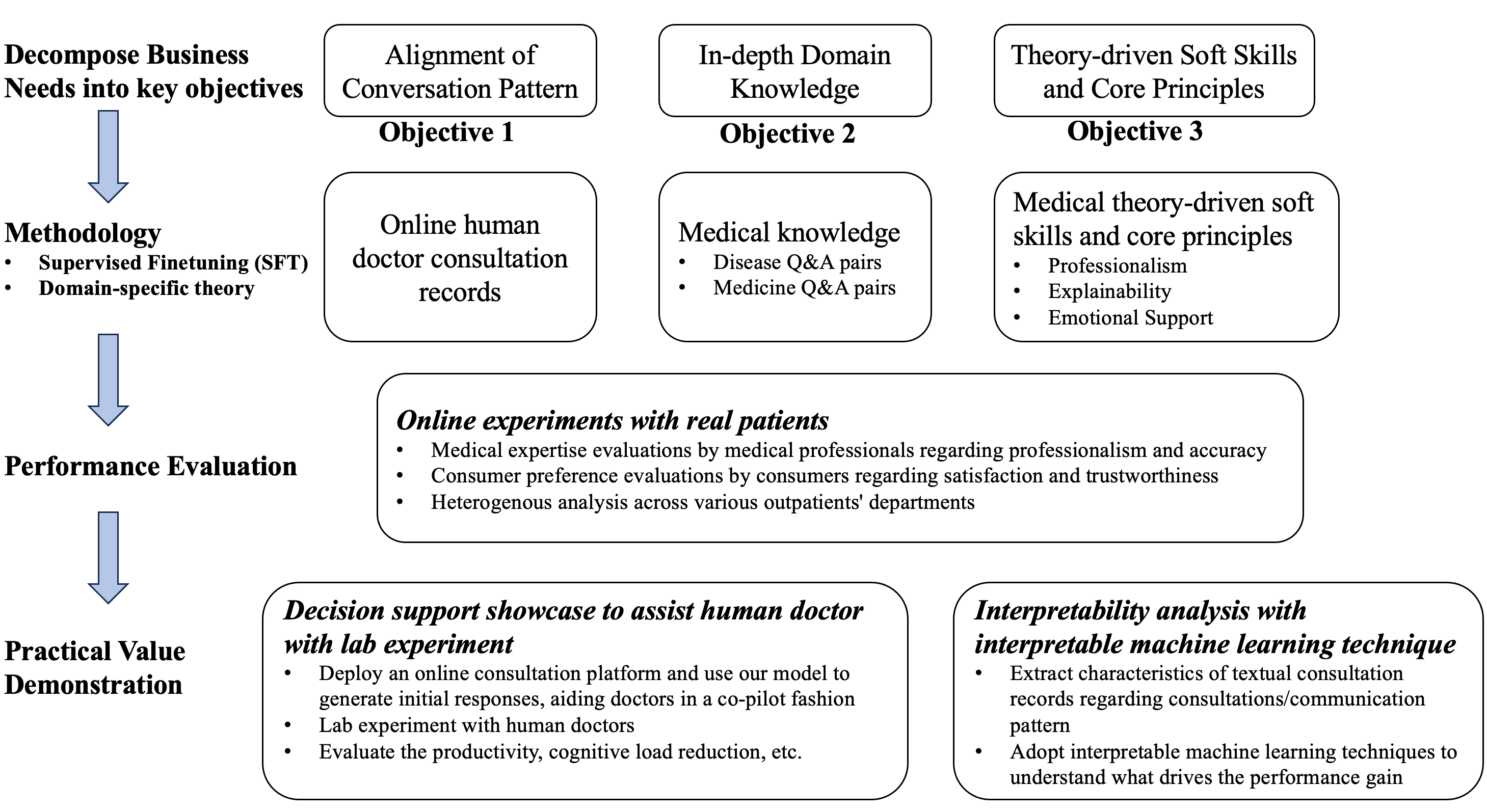}
\caption{Proposed Framework for Customizing LLMs for Business Contexts: A Medical Consultation Example}
\label{fig:framwork}
\end{figure}

The overall flow of the proposed framework is shown in Figure~\ref{fig:framwork}.

\subsection{Step 1: Construct Supervised Finetuning Dataset}



In this part, we  introduce the decomposed three fundamental objectives which represent the major  gaps between LLMs and a particular business context.  We also show how to achieve the objectives by compiling specific Supervised Fine Tuning data. 


$(i)$ \textbf{Objective 1: Alignment of conversational patterns of professional roles}: 
Each professional role adheres to specific communication structures and norms, reflecting its underlying business needs and domain-specific techniques. However, existing LLMs often deviate from these professional conversational patterns. To address this gap, it is essential to compile a substantial corpus of real-world professional conversation records, which inherently encapsulate these communication norms. Rather than establishing predefined rules, we can capitalize on the ability of LLMs to absorb vast amounts of data, enabling it to automatically understand the patterns from these large-scale authentic professional records. This approach allows for a more organic and accurate adaptation of LLMs to professional communication styles.

In the medical consultation context, we have gathered a large-scale collection of real doctors' online consultation records from Chunyu Doctor Ltd., a leading medical consultation platform in China. Chunyu Doctor is at the forefront of mobile internet healthcare and is a notable entity in the ``Internet + Healthcare'' sector. As of May 2022, it has garnered 150 million users and collaborated with over 660,000 practicing doctors from public hospitals. The platform has facilitated over 400 million patient services and compiled data on more than 300 million health profiles. With an average of over 390,000 daily consultations and a customer satisfaction rate of 98\%, Chunyu Doctor's extensive database provides an invaluable resource for LLMs to learn the conversational patterns of human doctors~\footnote{The statistics come from the official website:  \url{https://www.chunyuyisheng.com/about_us/}}.


We have collected more than 1.5 million consultation records. Each online consultation record covers multi-round patient-doctor conversation, which typically includes communication about the condition, diagnosis, and medicine instructions, etc. Specifically, in a standard medical consultation, a three-step diagnostic process is typically followed. Initially, the doctors engages in information collection, probing the patient for detailed information to better understand potential health concerns. This progresses to the diagnosis phase, where based on the information, a tentative diagnosis is drawn and accompanied by initial guidance or advice. Finally, in the treatment suggestion phase, the doctor outlines targeted treatment strategies suitable for the diagnosed condition, ensuring the patient is well-informed and prepared for the next steps in their healthcare journey.




$(ii)$ \textbf{Objective 2: In-depth domain knowledge}: 
Implementing LLMs in specific business domains requires a high level of domain-specific knowledge to ensure the accuracy and reliability of the provided information and advice. This entails a comprehensive understanding of the technical or specialized knowledge pertinent to the field. To meet this requirement, it is crucial to compile an exhaustive collection of domain knowledge, encompassing all possible information within the field, and present it in the format of question-answer pairs. This approach ensures that LLM is equipped with the necessary expertise to effectively operate in the specific business domain.

In the medical consultation domain, we construct a large-scale collection of medical knowledge Q\&A pairs, covering both disease and medicine knowledge. For disease knowledge, we source comprehensive data from Dingxiang Doctor, a leading medical information platform in China. Dingxiang Doctor is known for providing health consultation and science popularization services to the public, focusing on educational information, paid knowledge services, and healthcare e-commerce. The collected data includes detailed information on various aspects of diseases, such as symptoms, causes, diagnosis, treatment, lifestyle advice, prevention, and guidelines for consulting a doctor. This leads to  88,449 Q\&A pairs of diseases knowledge. 

\begin{figure}[tbh!]
\centering
\includegraphics[scale = 0.55]{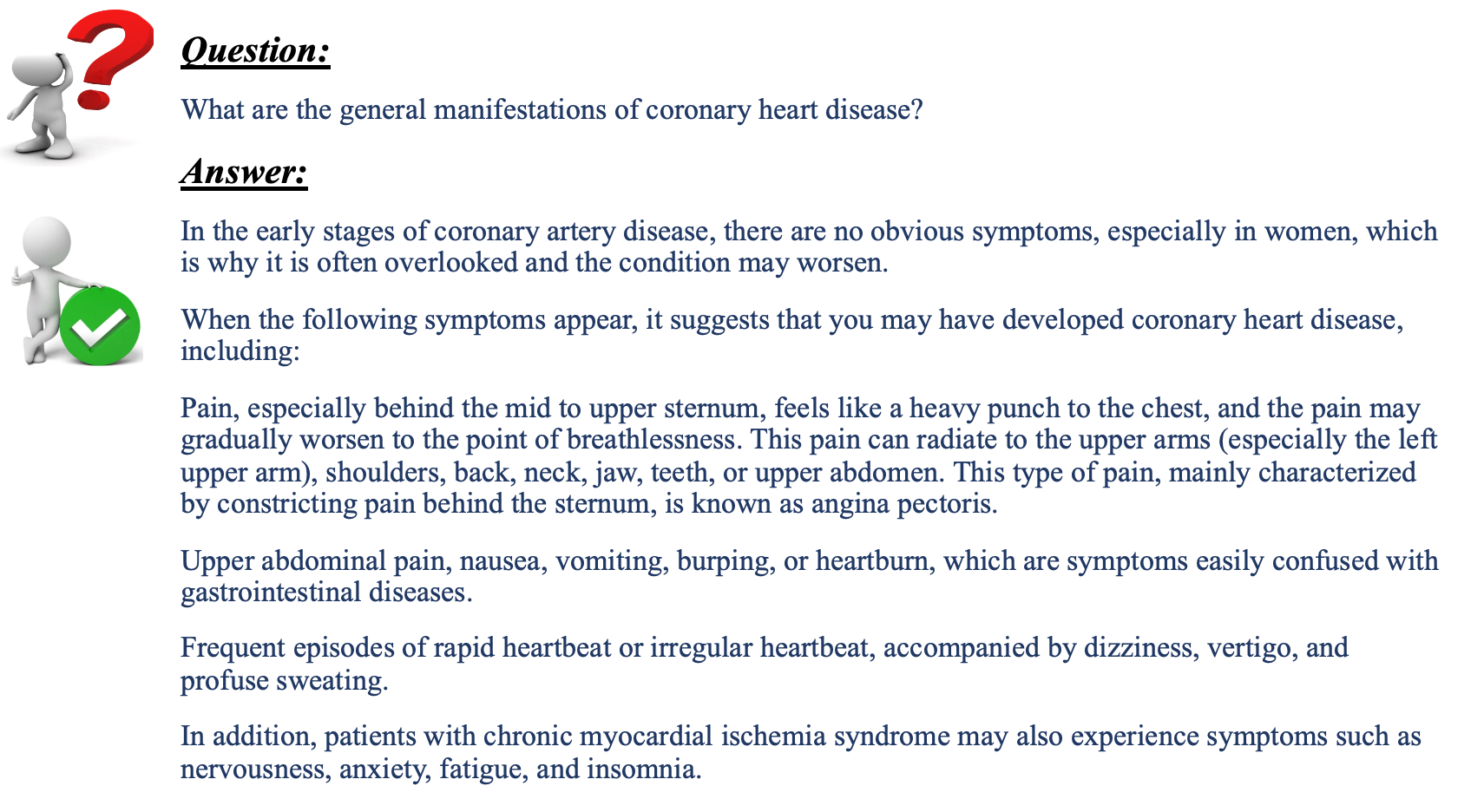}
\caption{Example of Disease Question Answer Pair}
\label{fig:diease_example}
\end{figure}

\begin{figure}[tbh!]
\centering
\includegraphics[scale = 0.55]{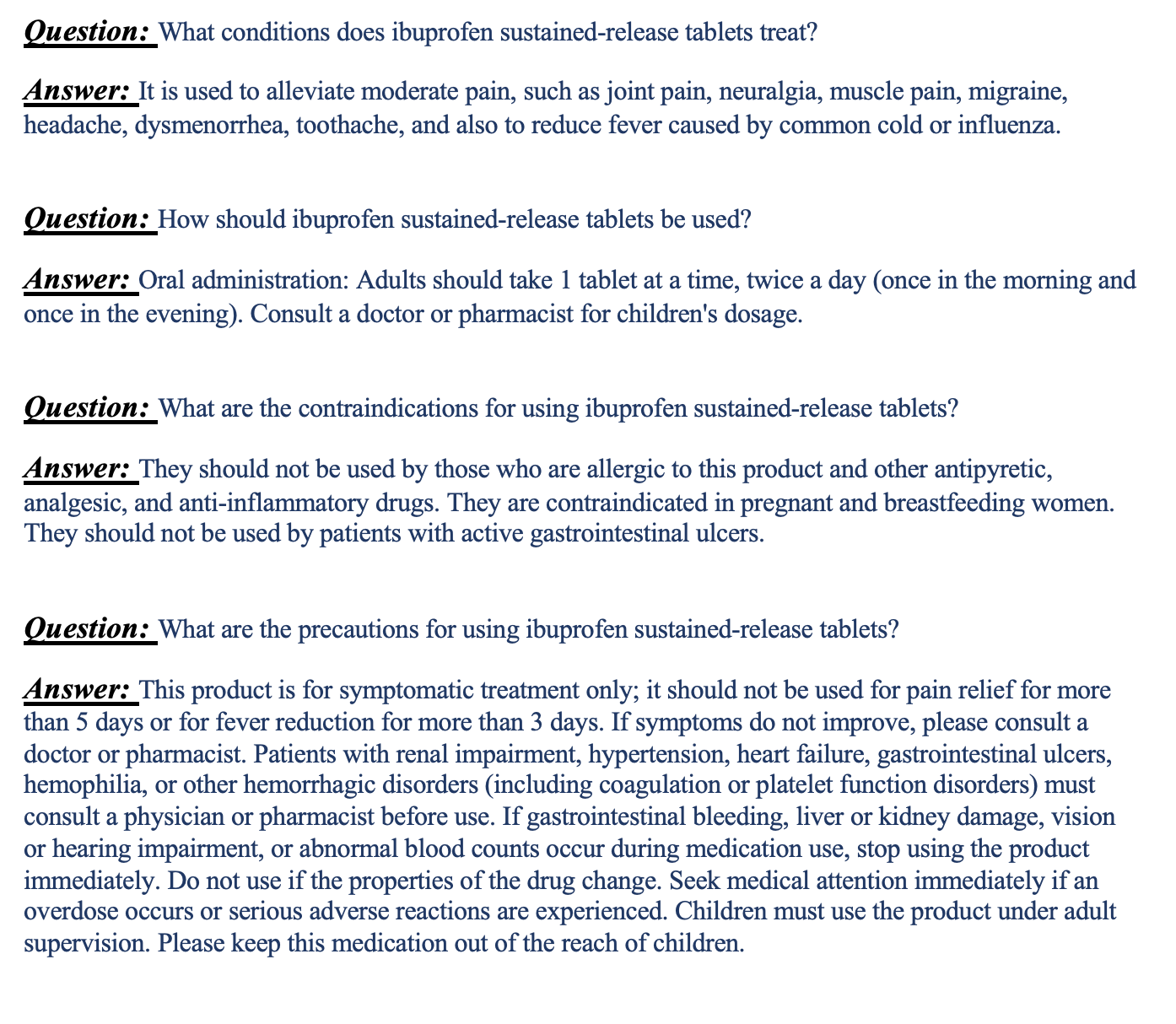}
\caption{Example of Medicine Question Answer Pair}
\label{fig:medicine_example}
\end{figure}

In terms of medicine knowledge, our source is Menet, a professional database specializing in medical and health industry research. It offers insights into hospital and retail markets, commercial channels, and internet-based medical information. We gather comprehensive details on 23,513 unique medicines from Menet. Using these medicines' instruction manuals, we create Q\&A pairs that cover various aspects such as usage, dosage, indications, contraindications, precautions, pharmacological effects, and chemical components. Questions are framed like ``What diseases does [Medicine Name] treat?'' for indications, or ``How is [Medicine Name] used?'' for dosage and administration. This  results in 88,163 Q\&A pairs, ensuring extensive coverage of each medicine's characteristics. Examples of disease and medicine Q\&A pairs are illustrated in Figures~\ref{fig:diease_example} and ~\ref{fig:medicine_example}, respectively.

$(ii)$ \textbf{Objective 3: Soft skills and core principles of professional roles}: 
For LLMs to effectively emulate professionals in various fields, it's crucial that it embodies soft skills and core principles akin to those of human professionals. Each business domain has its unique set of non-technical skills, interpersonal competencies, and foundational principles, essential for professional excellence. To achive this goal, our approach involves integrating domain-specific theories to identify the essential skills that professionals in each field should possess. We then use GPT-4 to evaluate conversation records along these well-defined dimensions. The goal is to pinpoint exemplary conversation records within large-scale datasets. These exemplary records inherently encode these soft skills, enabling LLMs to demonstrate similar skills and principles as  human professionals.

In the context of medical consultation, we demonstrate this process by first identifying the relevant soft skills and principles for human doctors from medical theory. Then, we select exemplary records that embody these identified skills.

\textbf{Medical Theoretical Foundations}:   According to medical theory, we identify three soft skills and core principles that a doctor should
follow include professionalism, explainability, and emotional support. 
These skills, when
combined, ensure not only patient safety but also cultivate trust,
ultimately contributing to a positive and reassuring healthcare experience for patients.



\begin{enumerate}

\item  \textit{Professionalism}: 
It has been identified as central to the practice of medicine, and its importance in healthcare has been widely discussed in the literature~\citep{wilkinson2009blueprint,kanter2013does,passi2010developing,cohen2006professionalism,van2009professionalism,wear2000development}. Professionalism in healthcare is the combination of ethical conduct, commitment to expertise, and interpersonal skills that uphold the trust society places in medical professionals, especially doctors. It entails a commitment to ethical practices such as honesty, integrity, and adherence to moral principles and professional codes~\citep{hilton2005proto,jha2006perceptions,swick2000toward,van2004conceptualize}. This professionalism requires effective communication with patients and their support networks, reliability through accountability, task completion, organization, and punctuality~\citep{frohna2005nature,project2002medical,kearney2005defining,wagner2007defining}. It also demands a dedication to ongoing self-improvement, lifelong learning, knowledge advancement, and fostering the development of others through feedback and leadership~\citep{hilton2005proto,jha2006perceptions,rabinowitz2004development,swick2000toward}.

Professionalism plays a pivotal role in medical consultations, particularly in the realm of patient-doctor communication, due to several compelling reasons. Firstly, it establishes a foundation of trust, which is essential for effective healthcare delivery. When doctors communicate with professionalism, marked by respect, empathy, and clarity, it fosters a safe environment where patients feel comfortable sharing sensitive information, crucial for accurate diagnoses and treatment plans. Secondly, professionalism encompasses the ability to communicate complex medical information in an accessible manner, ensuring that patients are well-informed about their health conditions and treatment options. This level of understanding is critical for patient engagement and adherence to medical advice. Moreover, professionalism entails adherence to ethical standards, including maintaining confidentiality, which is vital in preserving patient dignity and trust. Lastly, professional communication skills include the ability to listen actively and respond to patient concerns, demonstrating compassion and understanding. This not only enhances the therapeutic relationship but also directly impacts patient satisfaction and health outcomes. In summary, professionalism in patient-doctor communication is a key driver in ensuring effective, ethical, and patient-centered medical care.


\item \textit{Explainability}: Explainability of a doctor's communication is key to a successful patient–physician relationship and quality of care~\citep{ha2010doctor,markides2011importance}. It refers to the clarity and comprehensibility of the information conveyed by healthcare professionals to their patients. This involves simplifying complex medical terms and concepts into language that is easily understood, ensuring that patients have a solid understanding of their health conditions, treatment options, potential risks, and the benefits of the care they are receiving~\citep{hagihara2007association,kee2018communication,olson2010communication,freeman2019communicate,kee2018communication}. This concept is rooted in the principles of patient-centered care, which prioritize the patient’s comprehension, participation, and informed decision-making in their own healthcare.

 Explainability of a doctor's communication is paramount in healthcare as it directly influences patient outcomes. A physician who fosters transparent dialogue is more likely to gather comprehensive information, leading to accurate diagnoses and effective counseling. This enhances treatment adherence and promotes better long-term health. Conversely, insufficient explanations can result in misunderstanding, creating a disconnect between doctor and patient, which may cause treatment to fail~\citep{dimatteo1998role}. An efficient exchange of information is critical, ensuring patient concerns are addressed and treatment options are clearly explained, thereby facilitating shared decision-making~\citep{arora2003interacting,lee2002enhancing,kindler2005quantitative,minhas2007does}. Moreover, quality doctor-patient communication impacts various aspects of patient care, including satisfaction, treatment compliance, understanding of medical information, disease management, quality of life, and overall health~\citep{ong1995doctor,smith1981characteristics,larsen1981assessment,carter1982outcome}. Therefore, practices such as thorough explanations and active listening to patients or their families are essential in reducing adverse events~\citep{hagihara2007association}.


\item  \textit{Emotional Support}: Emotional support from doctors is a fundamental facet of patient care that substantially influences patient outcomes and satisfaction~\citep{bradshaw2022kindness,walter2021partnership,delbanco1992enriching}. This support extends beyond biomedical expertise to encompass essential psychosocial competencies~\citep{ommen2011relationship}. Patients share their hopes, fears, and concerns, seeking an empathetic ear and understanding~\citep{finset2012worried}. When doctors provide care that integrates clinical acumen with emotional support, the patient experience is markedly improved~\citep{han2019biopsychosocial,allen2018nature,northcott2018ve,rathert2015patient}. Emotional support in healthcare comprises cognitive, affective, and altruistic dimensions that collectively work to comprehend, empathize with, and ease patient distress~\citep{bivins2017compassionate,sharp2016vital}. Narrative knowing, which involves a mutual understanding between patient and doctor about the experience of illness, further deepens this support~\citep{buckley2018working}. Additionally, there are other practices to facilitate emotional support such as active listening and empathetic communication~\citep{bivins2017compassionate, babaei2019compassionate}.

Emotional support has been shown to enhance patients' engagement and satisfaction within online health communities, to a greater extent than informational support alone~\citep{chen2020exploring}. It can alleviate fears and anxieties, leading to higher levels of trust in doctors~\citep{delbanco1992enriching, cao2017doctors}.
It has been shown  emotional support is crucial for cancer patients and those nearing the end of life, offering solace and companionship during trying times~\citep{slevin1996emotional,wenrich2003dying}.
\end{enumerate}

\textbf{GPT rating:}
After identifying the relevant soft skills from medical theory, we employ a large-scale analysis of professional records to pinpoint exemplary cases demonstrating these skills. Utilizing GPT-4, we assess each online consultation record across these identified soft skill dimensions, rating them on a scale from 1 to 100. This process involves instructing GPT-4 to evaluate the performance of professionals (e.g., doctors) in specific areas, with a scoring range from 0 (indicating poor performance) to 100 (indicating excellent performance).



\begin{figure}[tbh!]
\centering
\includegraphics[scale = 0.8]{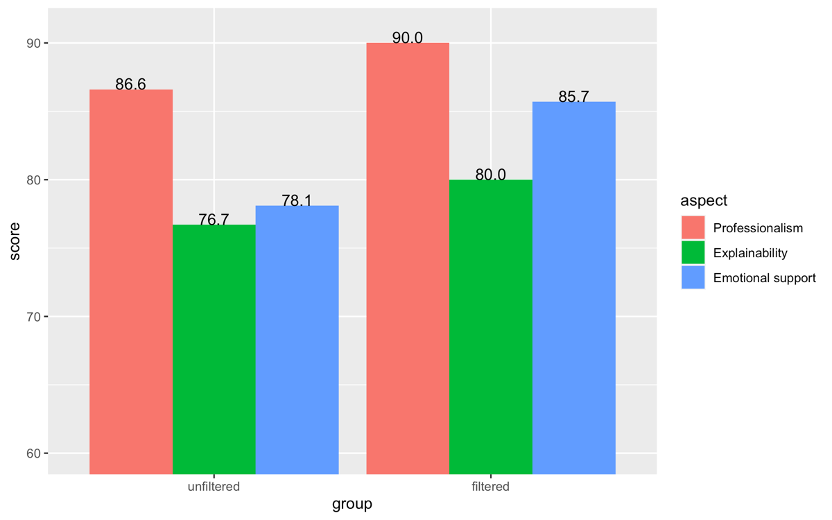}
\caption{Identify Role Model Records Based on Theory-driven Soft Skills and Core Principles of Human Doctors. All Dimensions Has Been Significantly Improved After Filter ($p < 0.001$)  }
\label{fig:theory_attributes}
\end{figure}

Using GPT-4's evaluations, we select the upper 50\% of records based on three essential skills, resulting in a total of 262,482 conversations. The distribution of these skills, both before and after this selection, is illustrated in Figure~\ref{fig:theory_attributes}, demonstrating a significant improvement in all three skills post-selection ($p < 0.001$). By using these role model medical records, we aim to make our LLM-doctor's  emulate the skills exhibited by these exemplary human doctors.


\subsection{Step 2: Finetuning Implementation}

After SFT data construction, we  perform supervised fine-tuning with open-source LLMs, specifically targeting the Baichuan2-13B-Chat model~\citep{baichuan2023baichuan2}. This model, developed by Baichuan Intelligence Inc., is a prominent open-source LLM for Chinese, fitting our dataset's language. Notably, our method is adaptable to other open-source LLMs. Baichuan2 is part of the latest generation of large-scale open-source language models, trained on a 2.6 trillion token corpus. It showcases exceptional performance in both Chinese and English benchmarks and is available in 7B and 13B versions for Base and Chat models, including a 4-bit quantized version for the Chat model. It has been thoroughly tested in six domains: general, legal, medical, mathematics, coding, and multilingual translation, demonstrating excellent Chinese-English translation capabilities and versatility in other languages.

For fine-tuning, we adopt the LoRA (Low-Rank Adaptation) strategy~\citep{hu2021lora}. This method adapts large pre-trained language models like GPT-3 without retraining all their parameters. LoRA differs from full fine-tuning by freezing the original model weights and introducing trainable rank decomposition matrices to each layer of the Transformer architecture. This method significantly reduces the number of trainable parameters and GPU memory requirements. LoRA has shown similar or even superior performance compared to traditional fine-tuning on models like RoBERTa, DeBERTa, GPT-2, and GPT-3, offering the benefits of greater training efficiency and no additional inference latency. 

Our training setup includes 8 Nvidia 100 GPUs, completing the training in 60 hours over four epochs. The hyperparameters are a global batch size of 16, a learning rate of 2e-5 using the AdamW optimizer, and a maximum sequence length of 1024 tokens.

\section{Model Performance Evaluation with Online Experiment}
\subsection{ Online Experiment}

We conduct  online experiments to evaluate the performance of our framework. We compare LLM-doctor with two benchmarks: one is the untuned base model, and the other is a human doctor. Such a comparison allows us to see how our framework improves over the base model as well as the gap with the human doctor.

We develop an online consultation platform that embeds either the LLM-doctor or an untuned base model in the backend to automatically generate responses for patients' inquiries and serve as a virtual doctor. We distribute this platform through the Credamo platform in China to recruit experimental participants. Credamo is similar to Amazon Mechanical Turk, facilitating large-scale data collection by connecting researchers with participants. We specifically recruit a large number of online patients who recently felt ill and needed medical consultation. Initially, we ask them if they were feeling unwell and required a doctor's consultation. If they responded affirmatively, we present them with a conversation interface where either the LLM-doctor or the untuned base model is embedded in the backend; otherwise, they are not eligible to participate in our study. This method help us filter for genuine patients to interact with our models. We recruit 1000 patients to interact with the LLM-doctor and another 1000 patients to engage with the base model. All consultation records, including patients' queries and the models' responses, are saved in real-time to our backend database on our platform.

To avoid perception bias, we introduce our service as a new online doctor platform without revealing its AI nature at the beginning. This creates an environment where users believe they are interacting with real doctors. 
This ensures a fair comparison with human doctor records without the concern of perception bias towards AI or humans. Our system is designed as a general-purpose medical consultation assistant, capable of addressing a broad range of medical inquiries without any disease-specific restrictions. Therefore, we do not set any constraints on acceptable disease categories. Instead, we welcome all inquiries from patients. Patients initiate conversations based on their unique health concerns. 

It's worth noting that our primary objective is to evaluate the effectiveness of our AI model and demonstrate its proof of concept. To ensure participants' safety and avoid potential harm, we display a message at the conclusion of their consultation—on the thank you page—clarifying that the responses were generated by an AI model, not a human doctor. Additionally, we specifically advise them not to consider the consultation as professional medical advice and to consult a human doctor for genuine medical guidance.

Following the collection of 2000 consultation records for LLM-doctor and the untuned base model, we employ GPT-4 to categorize each record into the appropriate broad outpatient departments, including internal medicine, surgery, head, neck, and vision, as well as other departments. The distribution of records is summarized in Table~\ref{table:sample_distribution}. We find that both LLM-doctor and the base model have a similar distribution over four outpatient departments, where the base model covers 40.4\% internal medicine, 25.1\% surgery, 13.6\% head, neck, and vision, and 20.9\% other departments, while LLM-doctor covers 43.1\% internal medicine, 22.1\% surgery, 14.9\% head, neck, and vision, and 19.9\% other departments. 
The proportion of each department is similar, and the coverage of the four departments is also similar. The relatively same distribution is important as it ensures that the comparison of the two models does not have a systematic difference in outpatient department coverage and ensures a fair comparison between the two models. Again, our system can handle all kinds of diseases; we don't set any constraints about disease or department. Patients initiate conversations based on their unique health concerns. Such distribution reflects the true medical demands from online patients/participants.

Additionally, we also want to compare with human doctors. We acknowledge the challenge of engaging a large number of human doctors. Therefore, we utilize the pre-collected real patient-doctor consultation records from Chunyu Doctor, which consist of actual responses from doctors to patient inquiries. To ensure a fair comparison with the base model and LLM-doctor, we randomly draw 1000 samples based on LLM-doctor's distribution over four departments. The sampled human doctor records distribution is summarized in Table~\ref{table:sample_distribution}. It has a relatively same distribution as the base model and LLM-doctor, covering 43.1\% internal medicine, 22.3\% surgery, 14.9\% head, neck, and vision, and 19.7\% other departments. It reveals that all three groups span a wide range of outpatient departments with roughly equivalent coverage. Consequently, there is no systematic difference in outpatient department coverage, ensuring a fair comparison among the groups.




\begin{table}[tbh!]
\scriptsize
\centering
\begin{tabular}{lllll|lll}
\toprule
Department& Base Model & LLM-doctor & Human Doctor \\
\midrule
Internal Medicine&  404 (40.4\%) & 431 (43.1\%) & 431 (43.1\%)\\
\hline

Surgery          &  251 (25.1\%) & 221 (22.1\%) & 223  (22.3\%)\\
\hline
Head, Neck, and Vision &  136 (13.6\%)& 149 (14.9\%) & 149 (14.9\%) \\
\hline
Other  Departments & 209 (20.9\%) & 199 (19.9\%) & 197 
 (19.7\%)\\
\hline
Total & 1000 &  1000 & 1000\\
\bottomrule
\end{tabular}
\caption{
Consultation Records Distribution Across Multiple Outpatient Departments }
\label{table:sample_distribution}
\end{table}

\subsection{Performance Evaluation with Medical Professionals and Real Consumers}

Our evaluation covers two key dimensions: medical expertise and consumer preference. For medical expertise, we assess consultations' \textit{professionalism} and  \textit{accuracy}. For consumer preference, we focus on patient \textit{satisfaction} and \textit{trustworthiness}. Firstly, professionalism and  accuracy directly relate to the quality of medical advice provided; ensuring that the system adheres to professional consultation procedures and standards and delivers precise diagnostics is paramount to patient safety and effective treatment. Secondly, patient satisfaction and trustworthiness are essential for gauging how well the system meets patient needs and expectations, which are crucial for patient engagement and adherence to medical advice. This comprehensive evaluation helps in refining the AI's capabilities and ensuring its utility in real-world healthcare settings.

For the medical expertise aspect, we have engaged a large group of real doctors from prominent hospitals in China. Their extensive clinical experience is crucial for accurate evaluations. We compare the consultation records from the untuned base model, LLM-doctor, and real doctors. We ask them to rate professionalism and accuracy along a scale of 0-100, where 100 represents extremely high and 0 represents extremely low. Each record is assessed by at least three different experts from the same department, ensuring a robust evaluation. The final score for each consultation record is the average of these ratings.

Regarding consumer preference, we involve real consumers who assess the consultation records from a patient's perspective. This is conducted on the Credamo platform. Participants rate the records based on trustworthiness and satisfaction, as if they were the patients in the consultations. We ask them to rate trustworthiness and satisfaction along a scale of 0-100, where 100 represents extremely high and 0 represents extremely low. Each record is reviewed by at least three different consumers. The final score for each consultation record is the average of these ratings.



The evaluations from both medical professionals and consumers are summarized in Table~\ref{table:expert}. Specifically, we report the average scores across various dimensions and also include the statistical significance, with t-statistics and p-values, for comparisons between two models.
Overall medical expertise is an average of professionalism and accuracy, overall consumer preference is an average of consumer satisfaction and trustworthiness, and overall performance is an average of professionalism, accuracy, satisfaction, and trustworthiness.

From a medical expertise standpoint, LLM-doctor significantly outperforms the base model with an +11.68\% improvement in professionalism ($p < 0.001$) and a +17.43\% improvement in accuracy ($p < 0.001$). Meanwhile, the base model significantly lags behind human doctors by -13.08\% in professionalism ($p < 0.001$) and by -16.9\% in accuracy ($p < 0.001$). Our LLM-doctor more closely matches the medical expertise of human doctors, with marginally smaller gaps of -2.93\% in professionalism ($p = 0.004$) and -2.42\% in accuracy ($p = 0.019$). This indicates that our framework significantly narrows the gap between untuned LLMs and human doctors, bringing it closer to human-level performance. This result is particularly significant, as a single LLM-model can approach the capabilities of nearly 1000 human doctors across different outpatient departments. This highlights the potential of the LLM-model to function as a general medical assistant, serving many human doctors in real worlds. 

The same conclusion applies to consumer preference metrics. From a consumer satisfaction perspective, the base model performs -27.76\% worse than human doctors ($p < 0.001$), and from a consumer trustworthiness perspective, it is -27.16\% worse ($p < 0.001$). The gap in consumer preferences between the base and human doctors is much larger than that in medical expertise. This indicates that the untuned base model significantly lags behind when engaging with consumers in reality, highlighting a potential shortcoming. This also underscores the need for a framework to improve in this aspect. From this standpoint, LLM-doctor significantly outperforms the base model with a +28.36\% improvement in consumer satisfaction ($p < 0.001$) and a +27.56\% improvement in trustworthiness ($p < 0.001$). It narrows the gap to human doctors' medical expertise with smaller deficits of -7.27\% in satisfaction ($p < 0.001$) and -7.09\% in trustworthiness ($p < 0.001$). This further indicates that our framework significantly narrows the gap between untuned LLMs and human doctors, moving it closer to human-level performance, with an even stronger improvement from the perspective of consumer preference.

\begin{table}[tbh!]
\scriptsize
\centering
\begin{tabular}{llll|llll}
\toprule
&&&& LLM-doctor & LLM-doctor & Base \\
&Base&LLM&Human& vs& vs & vs\\
Evaluation metrics&   Model   & Doctor  &  Doctor & Base& Human Doctor & Human Doctor\\

\midrule
Professionalism& 67.78&	75.70&	77.98 &  +11.68\% & -2.93\%  & -13.08\%\\
 & & & & (9.29, $p < 0.001$) & (-2.86, $p = 0.004$) & (-15.45, $p < 0.001$) \\
Accuracy&   63.17&	74.18&	76.01&  +17.43\% & -2.42\% & -16.90\%\\
 & & & &  (12.77, $p < 0.001$) & (-2.35, $p = 0.019$) & (-20.83, $p < 0.001$)               \\
\textit{Overall Medical Expertise} & 65.47&	74.94	&77.00& +14.45\% & -2.68\% & -14.97\%\\
 & & & & (11.36, $p < 0.001$) & (-2.70,	$p = 0.007$) & (-19.32, $p < 0.001$) \\
\hline
Satisfaction&    56.24&	72.19&	77.86 &  +28.36\% & -7.27\% & -27.76\%\\

 & & & &   (21.51, $p < 0.001$)& (-7.87,	$p < 0.001$)   & (-29.37, $p < 0.001$)        \\
Trustworthiness&     56.68&	72.30&	77.82& +27.56\% & -7.09\% & -27.16\% \\
 & & & &    (20.69, $p < 0.001$)  & (-7.57, $p < 0.001$) & (-28.32, $p < 0.001$)       \\
\textit{Overall Consumer Preference}& 56.46&	72.25&	77.84& +27.96\% & -7.18\% & -27.46\% \\
 & & & &   (21.41, $p < 0.001$) & (-7.81, $p < 0.001$) & (-29.26, $p < 0.001$)    \\
\hline
\textit{Overall Performance} & 60.97&	73.59&	77.42 & +20.71\% & -4.94\% & -21.25\%\\
 & & & &   (22.81, $p < 0.001$)& (-7.21, $p < 0.001$) & (-35.04, $p < 0.001$)         \\
Number of entities& 1 & 1 & 1000          \\
Number of records&1000 &  1000 & 1000         \\
\bottomrule
\end{tabular}
\caption{
Medical Experts and Real Consumers Evaluation: Model Performance Comparison  for  Base Model, LLM-doctor and Human Doctors Along Medical Expertise Metrics and Consumer Preference Metrics.   }
\label{table:expert}
Note: We conduct two sample t-test and report the t-stats and p-value to compare two model. 
\end{table}

\subsection{Heterogeneous Comparison Across Various Outpatients Departments}

Beyond average performance, we delve into various outpatient departments and compare the nuanced performance in each department as shown in Table~\ref{table:hetergeneous_analysis}. Such analysis can help us understand how the performance of our framework differs across heterogeneous outpatient departments.

We find that in the internal medicine department, the LLM-doctor shows a relatively larger overall improvement (an average of medical expertise and consumer preference) compared to the untuned base model, about +21.64\% ($p < 0.001$), followed by the surgery department with an improvement of +16.76\% ($p < 0.001$). The smallest improvement occurs in the head, neck, and vision departments with an improvement of +15.82\% ($p < 0.001$). Correspondingly, in the internal medicine department, the LLM-doctor is closest to the human doctor with a  -4.74\% gap ($p < 0.001$), followed by a -5\% gap  ($p < 0.001$)in the surgery department. The largest gap, -6.5\% ($p < 0.001$), occurs in the head, neck, and vision departments. This indicates that our framework functions better in internal medicine and could be prioritized for use in this department compared to others.

\begin{table}[tbh!]
\scriptsize
\centering
\begin{tabular}{lllll|lll}
\toprule
&&&&& LLM-doctor & LLM-doctor & Base \\
&&Base&LLM&Human& vs& vs & vs\\
Department&Evaluation metrics&   Model   & Doctor  &  Doctor & Base& Human Doctor & Human Doctor\\

\midrule
\multirow{7}{*}{Internal Medicine} & Professionalism & 69.53 & 75.84 & 79.43 & +9.08\% & -4.52\% & -12.46\%	\\
 & & & & & (9.52), $p < 0.001$ & (-4.27), $p < 0.001$ & (-11.41), $p < 0.001$	\\
 & Accuracy & 64.60 & 74.73 & 76.98 & +15.68\% & -2.92\% & -16.07\%	\\
 & & & & & (13.79), $p < 0.001$ & (-3.42), $p < 0.001$ & (-18.38), $p < 0.001$	\\
 & Satisfaction & 54.85 & 72.99 & 77.55 & +33.07\% & -5.88\% & -29.27\%	\\
 & & & & & (16.91), $p < 0.001$ & (-4.22), $p < 0.001$ & (-19.82), $p < 0.001$	\\
 & Trustworthiness & 55.02 & 73.26 & 77.64 & +33.14\% & -5.64\% & -29.13\%	\\
 & & & & & (16.50), $p < 0.001$ & (-3.99), $p < 0.001$ & (-19.49), $p < 0.001$	\\
 & \textit{Medical Expertise} & 67.07 & 75.29 & 78.20 & +12.26\% & -3.73\% & -14.24\%	\\
 & & & & & (13.13), $p < 0.001$ & (-4.48), $p < 0.001$ & (-17.23), $p < 0.001$	\\
 & \textit{Consumer Preference} & 54.94 & 73.13 & 77.60 & +33.10\% & -5.76\% & -29.20\%	\\
 & & & & & (17.00), $p < 0.001$ & (-4.15), $p < 0.001$ & (-19.91), $p < 0.001$	\\
 & \textit{Overall Performance} & 61.00 & 74.21 & 77.90 & +21.64\% & -4.74\% & -21.69\%	\\
 & & & & & (21.16), $p < 0.001$ & (-5.61), $p < 0.001$ & (-25.79), $p < 0.001$	\\
\hline

\multirow{7}{*}{Surgery} & Professionalism & 71.95 & 76.38 & 77.44 & +6.15\% & -1.37\% & -7.09\%	\\
 & & & & & (5.84), $p < 0.001$ & (-1.30), $p = 0.194$ & (-7.00), $p < 0.001$	\\
 & Accuracy & 65.34 & 74.74 & 75.88 & +14.38\% & -1.50\% & -13.89\%	\\
 & & & & & (10.65), $p < 0.001$ & (-1.30), $p = 0.194$ & (-11.88), $p < 0.001$	\\
 & \textit{Medical Expertise} & 68.65 & 75.56 & 76.66 & +10.07\% & -1.44\% & -10.45\%	\\
 & & & & & (9.51), $p < 0.001$ & (-1.36), $p = 0.174$ & (-10.57), $p < 0.001$	\\
 & Satisfaction & 57.38 & 71.71 & 78.53 & +24.98\% & -8.68\% & -26.94\%	\\
 & & & & & (8.98), $p < 0.001$ & (-4.38), $p < 0.001$ & (-13.79), $p < 0.001$	\\
 & Trustworthiness & 57.79 & 71.94 & 78.43 & +24.49\% & -8.28\% & -26.32\%	\\
 & & & & & (8.85), $p < 0.001$ & (-4.22), $p < 0.001$ & (-13.29), $p < 0.001$	\\
 & \textit{Consumer Preference} & 57.58 & 71.82 & 78.48 & +24.73\% & -8.48\% & -26.63\%	\\
 & & & & & (9.01), $p < 0.001$ & (-4.35), $p < 0.001$ & (-13.72), $p < 0.001$	\\
 & \textit{Overall Performance} & 63.11 & 73.69 & 77.57 & +16.76\% & -5.00\% & -18.64\%	\\
 & & & & & (11.88), $p < 0.001$ & (-4.36), $p < 0.001$ & (-17.17), $p < 0.001$	\\
\hline

\multirow{7}{*}{Head, Neck, and Vision} & Professionalism & 70.11 & 75.17 & 77.58 & +7.21\% & -3.11\% & -9.63\%	\\
 & & & & & (5.60), $p < 0.001$ & (-2.74), $p = 0.006$ & (-8.65), $p < 0.001$	\\
 & Accuracy & 66.58 & 73.71 & 75.34 & +10.71\% & -2.16\% & -11.63\%	\\
 & & & & & (6.88), $p < 0.001$ & (-1.65), $p = 0.100$ & (-8.49), $p < 0.001$	\\
 & \textit{Medical Expertise} & 68.35 & 74.44 & 76.46 & +8.92\% & -2.64\% & -10.62\%	\\
 & & & & & (7.30), $p < 0.001$ & (-2.30), $p = 0.022$ & (-9.78), $p < 0.001$	\\
 & Satisfaction & 55.70 & 70.14 & 78.15 & +25.93\% & -10.25\% & -28.73\%	\\
 & & & & & (6.35), $p < 0.001$ & (-4.01), $p < 0.001$ & (-11.95), $p < 0.001$	\\
 & Trustworthiness & 57.12 & 69.97 & 78.01 & +22.49\% & -10.31\% & -26.78\%	\\
 & & & & & (5.59), $p < 0.001$ & (-3.98), $p < 0.001$ & (-11.18), $p < 0.001$	\\
 & \textit{Consumer Preference} & 56.41 & 70.05 & 78.08 & +24.19\% & -10.28\% & -27.75\%	\\
 & & & & & (6.03), $p < 0.001$ & (-4.04), $p < 0.001$ & (-11.74), $p < 0.001$	\\
 & \textit{Overall Performance} & 62.38 & 72.25 & 77.27 & +15.82\% & -6.50\% & -19.27\%	\\
 & & & & & (8.09), $p < 0.001$ & (-4.45), $p < 0.001$ & (-13.92), $p < 0.001$	\\
\hline

\multirow{7}{*}{Other  Departments} & Professionalism & 57.88 & 75.02 & 75.73 & +29.62\% & -0.93\% & -23.57\%	\\
 & & & & & (4.60), $p < 0.001$ & (-0.21), $p = 0.836$ & (-7.59), $p < 0.001$	\\
 & Accuracy & 55.56 & 72.70 & 74.57 & +30.85\% & -2.50\% & -25.49\%	\\
 & & & & & (4.63), $p < 0.001$ & (-0.54), $p = 0.590$ & (-8.22), $p < 0.001$	\\
 & \textit{Medical Expertise} & 56.72 & 73.86 & 75.15 & +30.22\% & -1.71\% & -24.52\%	\\
 & & & & & (4.65), $p < 0.001$ & (-0.38), $p = 0.706$ & (-8.04), $p < 0.001$	\\
 & Satisfaction & 57.91 & 72.54 & 77.53 & +25.25\% & -6.44\% & -25.30\%	\\
 & & & & & (9.31), $p < 0.001$ & (-3.24), $p = 0.001$ & (-12.02), $p < 0.001$	\\
 & Trustworthiness & 58.26 & 72.36 & 77.38 & +24.19\% & -6.48\% & -24.70\%	\\
 & & & & & (8.80), $p < 0.001$ & (-3.12), $p = 0.002$ & (-11.37), $p < 0.001$	\\
 & \textit{Consumer Preference} & 58.09 & 72.45 & 77.45 & +24.72\% & -6.46\% & -25.00\%	\\
 & & & & & (9.22), $p < 0.001$ & (-3.23), $p = 0.001$ & (-11.90), $p < 0.001$	\\
 & \textit{Overall Performance} & 57.40 & 73.16 & 76.30 & +27.44\% & -4.12\% & -24.76\%	\\
 & & & & & (8.01), $p < 0.001$ & (-1.70), $p = 0.090$ & (-13.63), $p < 0.001$	\\

\hline
\bottomrule
\end{tabular}
\caption{
Model Comparison Along Heterogeneous Outpatient Departments  }
\label{table:hetergeneous_analysis}
Note: Note: We conduct two sample t-test and report the t-stats and p-value to compare two model. 
\end{table}

\section{Interpretability analysis: Understand Where the Gains Come From}

In the previous section, we compare the performance of three models and demonstrate the superiority of our framework in terms of both medical expertise and perceived consumer satisfaction and trustworthiness as determined by consumers. However, it is unclear what happens in conversation records and where the performance gain come from, particularly which conversation or communication characteristics contribute to the performance gains. To answer this question, we delve deep into the textual conversation records between patients and the LLM-doctor, base model, and human doctor, extracting textual communication and consultation characteristics from consultation records. Afterwards, we use an interpretable machine learning model, XGBoost, to understand how these extracted textual characteristics influence perceived medical expertise and consumer preference.

\subsection{Analyze Textual Communication and Consultation Characteristics}

\begin{figure}[!hbpt]
    \centering
    \includegraphics[scale = 0.3]{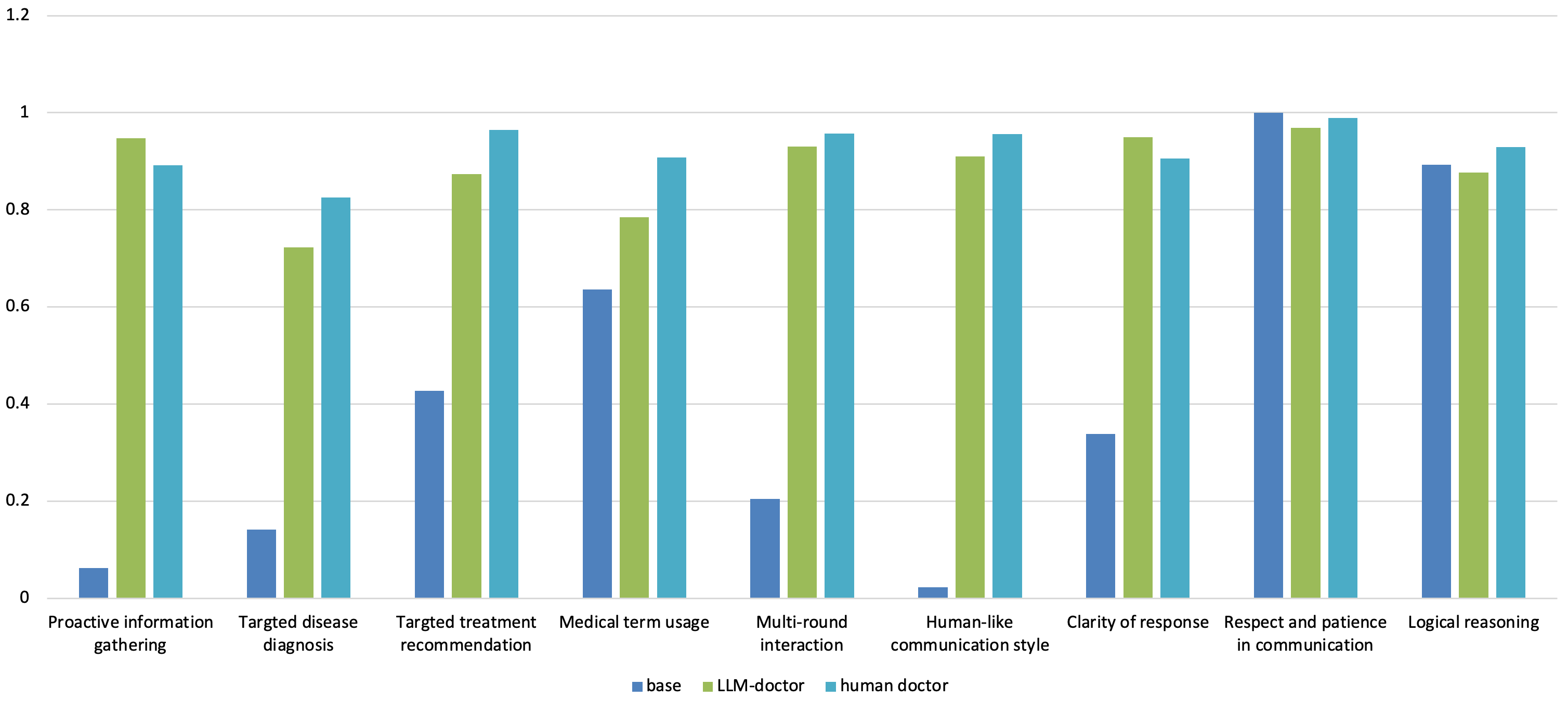}
    \caption{\centering Visualization of Medical Consultation Record Characteristics Comparison in Terms of Percentage of "Yes" Answers Across Characteristics}
    \label{textual_features}
\end{figure}

Given the proven reliability of GPT-4 in annotations and evaluations (as noted in sources such as Gilardi 2023, Eloundou 2023, and Lou 2023), we utilize GPT-4 to extract characteristics from medical records across various textual communication and consultation features. To bypass difficulties or ambiguities, we simplify the classification into a binary decision—yes or no—specifically determining whether the doctor's record demonstrates proactive information gathering, provides targeted disease diagnosis, offers targeted treatment recommendations, uses medical terms, engages in multi-round interactions, exhibits a human-like communication style, demonstrates clarity of response, and shows respect and patience in communication, as well as logical reasoning. Consider proactive information gathering as an example. The prompt for GPT-4 to do classification is: Below is a conversation between a patient and a doctor during an online consultation. 
Please determine if the doctor first asked questions to learn more about the patient's health condition and gathered more information before making any diagnosis and treatment recommendations (Yes), or if the doctor did not ask any further questions and made a direct diagnosis and treatment recommendations (No). Please answer with either "Yes" or "No," and only return the answer. We calculate the percentage of "Yes" answers for each model across each characteristic and plot the distribution in Figure~\ref{textual_features}, and also report the significance test for comparison between models across various textual characteristics in Table~\ref{table:texual_features_comparesion}.

As we observe, both LLM-doctor (94.7\% of consultation records) and human doctors (89.2\% of consultation records) consistently engage in proactive information gathering by asking patients for more details about their health conditions before making diagnoses and suggestions, whereas the base model seldom does with only 6.2\% of consultation records proactively asking. Proactive information gathering is extremely important for doctors to fully understand patients' health situations and provide precise diagnosis and recommendations. This suggests the base model significantly falls short in this manner. Our LLM-doctor is much closer to real doctors' consultation procedures. Additionally, both LLM-doctor (72.3\% of consultation records) and human doctors (82.6\% of consultation records) provide targeted diagnosis, whereas the base model rarely (only 14.1\% of consultation records) provides targeted responses and often includes many irrelevant details. Regarding targeted treatment recommendations, human doctors provide targeted treatment recommendations based on patient's health conditions in 96.5\% of all consultation records, LLM-doctor in 87.4\% of all consultation records, whereas the base model only provides 42.7\% targeted treatment recommendations. Regarding medical term usage, human doctors use medical terms in 90.8\% of cases during consultations, LLM-doctor in 78.5\% of cases, whereas the base model uses medical terms less often, only in 63.6\% of cases.

Furthermore, LLM-doctor (93\% of consultation records) and human doctors (95.7\% of consultation records) often engage in multi-round interactions, whereas only 20.5\% of base model consultation records have multi-round interactions. The majority of consultation records from both LLM-doctor and human doctors exhibit a human-like communication style, with human doctors showing this style in 95.6\% of consultations, and LLM-doctor in 91\% of records. However, the base model only demonstrates this style 2.3\% of the time. This suggests that the base model is far removed from human communication styles, resembling more of a bot-style communication. This underscores the benefit of aligning conversation patterns to make our model more akin to a human doctor. Additionally, the majority of human doctors' responses exhibit high clarity with 90.6\% of cases, LLM-doctor with 95\% of cases, whereas the base model's clarity is relatively low with only 33.8\% of cases exhibiting high clarity. This indicates the response of the base model lacks clarity, is very long, and includes redundant and irrelevant information.

To summarize, LLM-doctor is much closer to human doctors along the majority of these textual communication and consultation characteristics, whereas the base model is relatively far from human doctors. This indicates our framework can significantly change how LLMs conduct consultations such as the procedure, and communication styles, making it more human doctor-like.

\begin{table}[tbh!]
\scriptsize
\centering
\begin{tabular}{llll|llll}
\toprule
&&&& LLM-doctor & LLM-doctor & Base \\
&Base&LLM&Human& vs& vs & vs\\
Evaluation metrics&   Model   & Doctor  &  Doctor & Base& Human Doctor & Human Doctor\\
\midrule

proactive information gathering & 6.2\% & 94.7\%  & 89.2\%  & +1427.42\% & +6.17\% & -93.05\%	\\
 & & & & (84.98), $p < 0.001$ & (4.54), $p < 0.001$ & (-66.74), $p < 0.001$	\\
targeted disease diagnosis & 14.1\%  & 72.3\%  & 82.6\%  & +412.77\% & -12.47\% & -82.93\%	\\
 & & & & (32.45), $p < 0.001$ & (-5.55), $p < 0.001$ & (-42.07), $p < 0.001$	\\
targeted treatment recommendation & 42.7\%  & 87.4\%  & 96.5\%  & +104.68\% & -9.43\% & -55.75\%	\\
 & & & & (23.72), $p < 0.001$ & (-7.58), $p < 0.001$ & (-32.23), $p < 0.001$	\\
medical term usage & 63.6\%  & 78.5\%  & 90.8\%  & +23.43\% & -13.55\% & -29.96\%	\\
 & & & & (7.44), $p < 0.001$ & (-7.74), $p < 0.001$ & (-15.32), $p < 0.001$	\\
multi-round interaction & 20.5\%  & 93\%  & 95.7\%  & +353.66\% & -2.82\% & -78.58\%	\\
 & & & & (47.98), $p < 0.001$ & (-2.62), $p = 0.009$ & (-52.61), $p < 0.001$	\\
human-like communication style & 2.3\%  & 91\%  & 95.6\%  & +3856.52\% & -4.81\% & -97.59\%	\\
 & & & & (86.78), $p < 0.001$ & (-4.13), $p < 0.001$ & (-116.08), $p < 0.001$	\\
clarity of response & 33.8\%  & 95\%  & 90.6\%  & +181.07\% & +4.86\% & -62.69\%	\\
 & & & & (37.14), $p < 0.001$ & (3.82), $p < 0.001$ & (-32.30), $p < 0.001$	\\
respect and patience in communication & 100\%  & 96.9\%  & 98.9\%  & -3.10\% & -2.02\% & +1.11\%	\\
 & & & & (-5.65), $p < 0.001$ & (-3.13), $p = 0.002$ & (3.33), $p < 0.001$	\\
logical reasoning & 89.3\%  & 87.7\%  & 92.9\%  & -1.79\% & -5.60\% & -3.88\%	\\
 & & & & (-1.12), $p = 0.262$ & (-3.94), $p < 0.001$ & (-2.83), $p = 0.005$	\\

\bottomrule
\end{tabular}
\caption{
Medical Consultation Record Characteristics Comparison For Base Model, LLM-doctor and Human Doctors. We Show Percentage of ”Yes” Answers for Each Model across Each Characteristic }
\label{table:texual_features_comparesion}
Note: We conduct two sample t-test and report the t-stats and p-value to compare two model. 
\end{table}

\subsection{Interpretable Machine Learning Analysis}

We then employ the interpretable machine learning model to understand how these extracted textual communication and consultation characteristics influence perceived medical expertise and consumer preference. Specifically, we adopt XGBoost~\citep{chen2016xgboost}, which is a highly efficient and scalable implementation of gradient boosted trees, designed to push the limits of computing power for boosted tree algorithms. It has gained significant popularity in business applications due to its robust performance and its ability to provide interpretable insights through feature importance metrics~\citep{zhang2023can, wang2023learning}. This feature allows users to understand which variables most significantly impact the model's predictions, offering valuable insights into the underlying data patterns and decision-making processes.

\begin{figure}[!ht]
     \centering
     \begin{subfigure}[b]{0.8\textwidth }
         \centering
         \includegraphics[width=\textwidth, height=5.5cm]{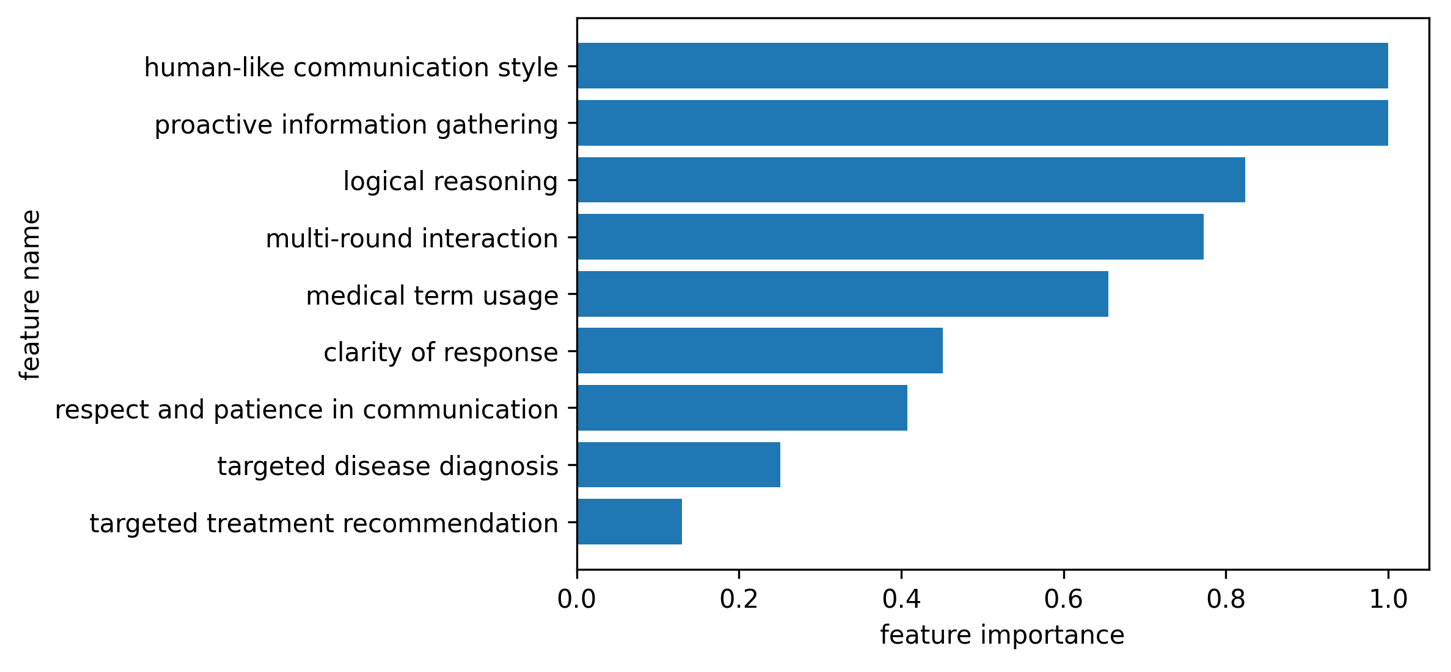}
         \caption{Professionalism}
     \end{subfigure}
     \hfill
     \begin{subfigure}[b]{0.8\textwidth}
         \centering
         \includegraphics[width=\textwidth, height=5.5cm]{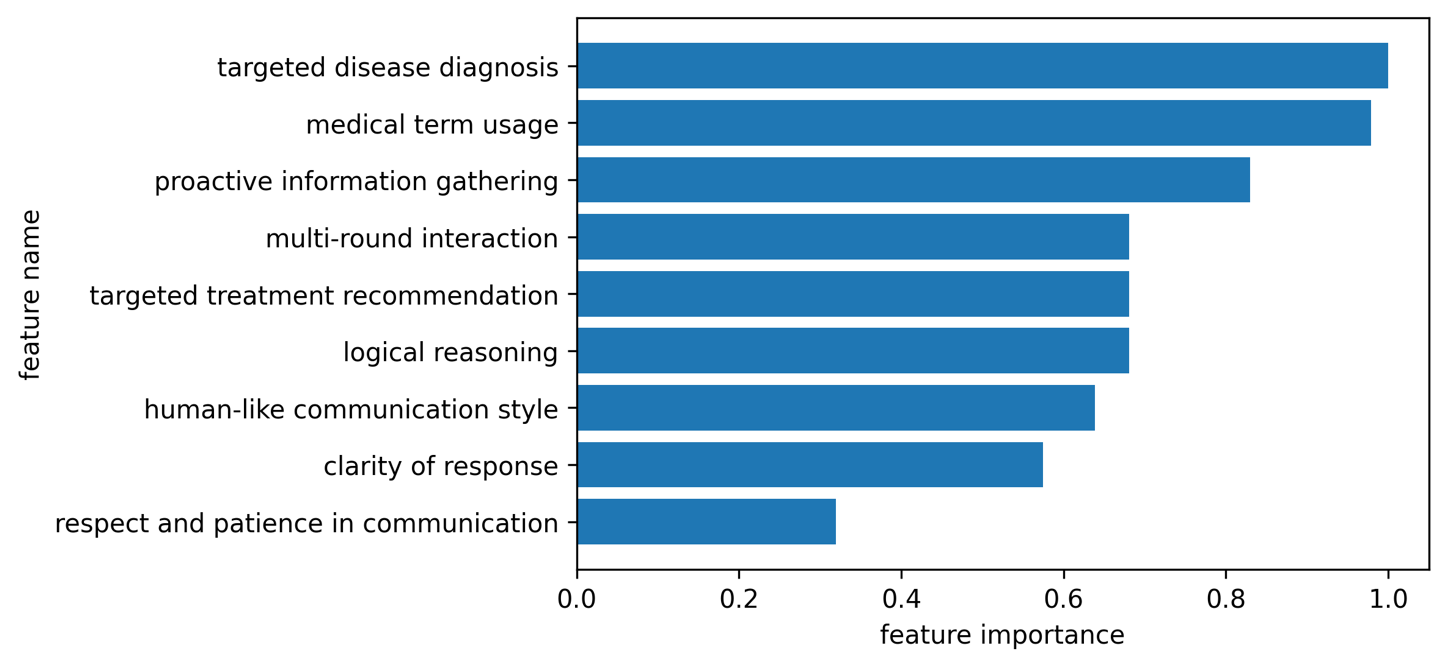}
         \caption{Accuracy}
     \end{subfigure}    
          \hfill
     \begin{subfigure}[b]{0.8\textwidth}
         \centering
         \includegraphics[width=\textwidth, height=5.5cm]{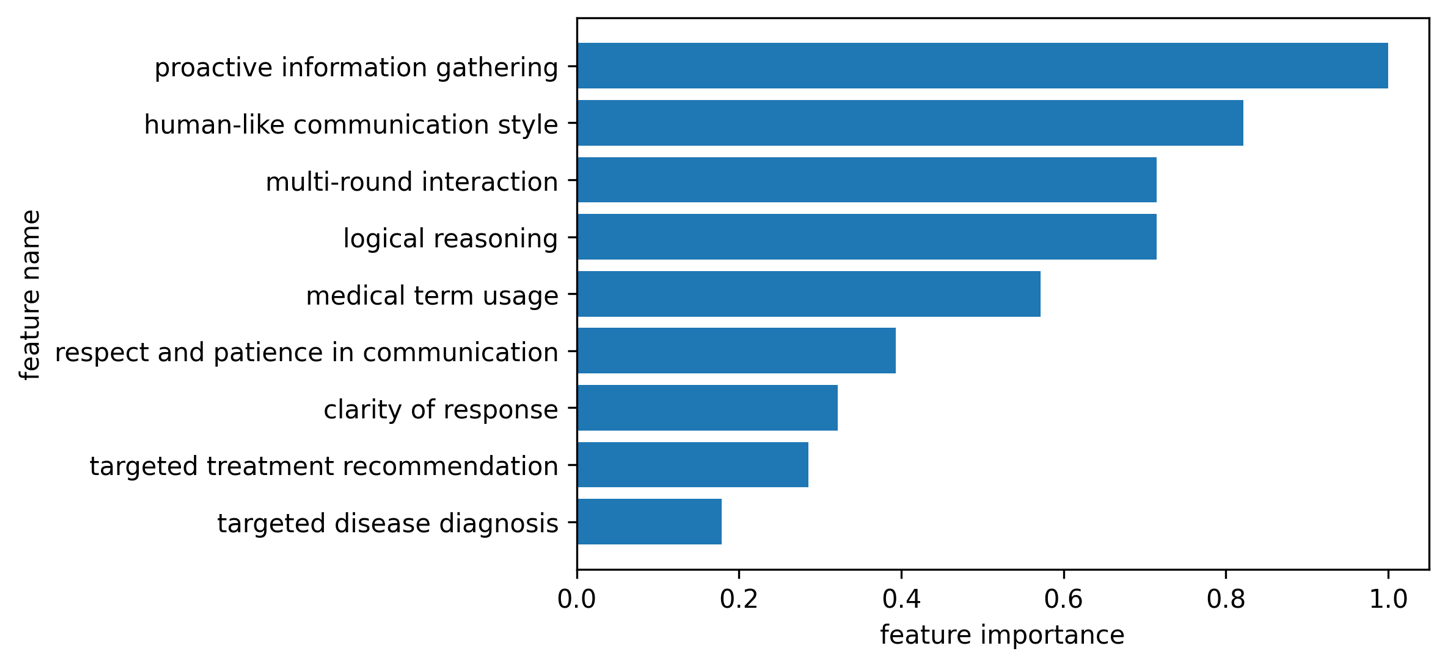}
         \caption{Overall Medical Expertise}
     \end{subfigure}    
        \caption{Feature Importance of Medical Consultation Record Characteristics on Medical Expertise} 
        \label{feature_importance_medical_expertise}
\end{figure}

\begin{figure}[!ht]
     \centering
     \begin{subfigure}[b]{0.8\textwidth }
         \centering
         \includegraphics[width=\textwidth, height=5.5cm]{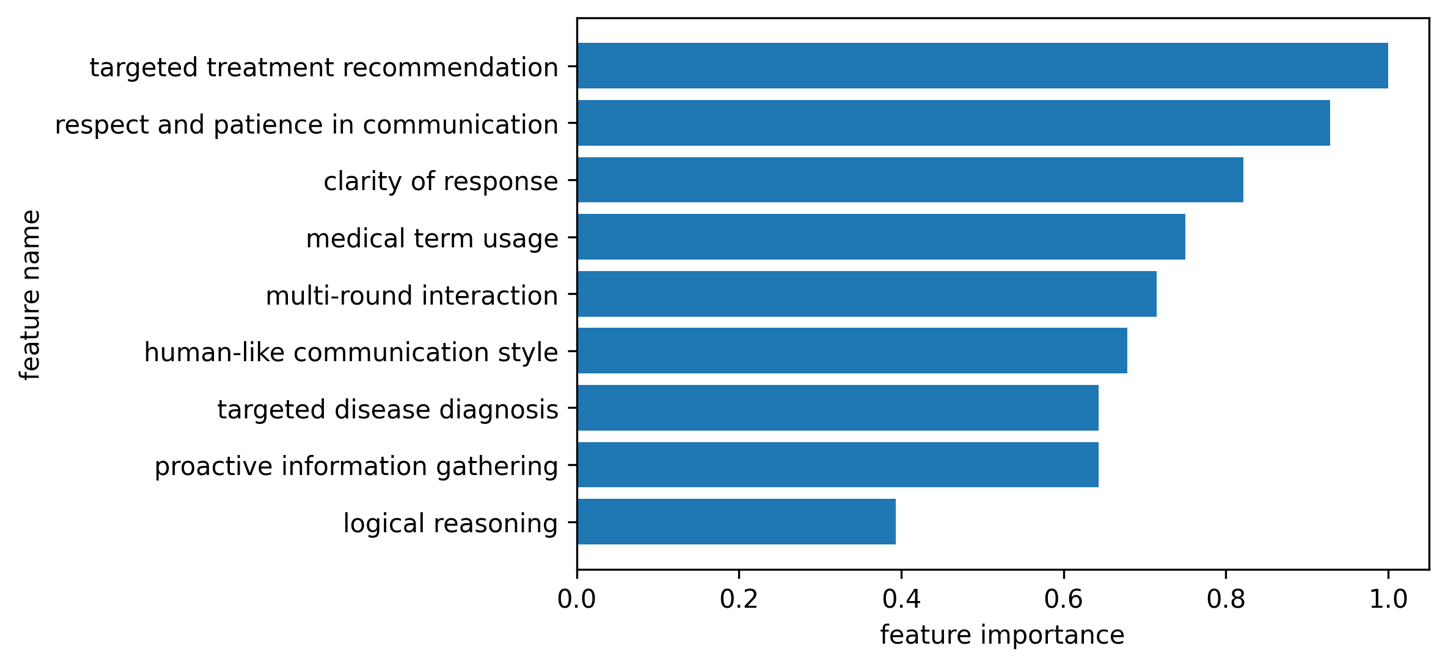}
         \caption{Satisfaction}
     \end{subfigure}
     \hfill
     \begin{subfigure}[b]{0.8\textwidth}
         \centering
         \includegraphics[width=\textwidth, height=5.5cm]{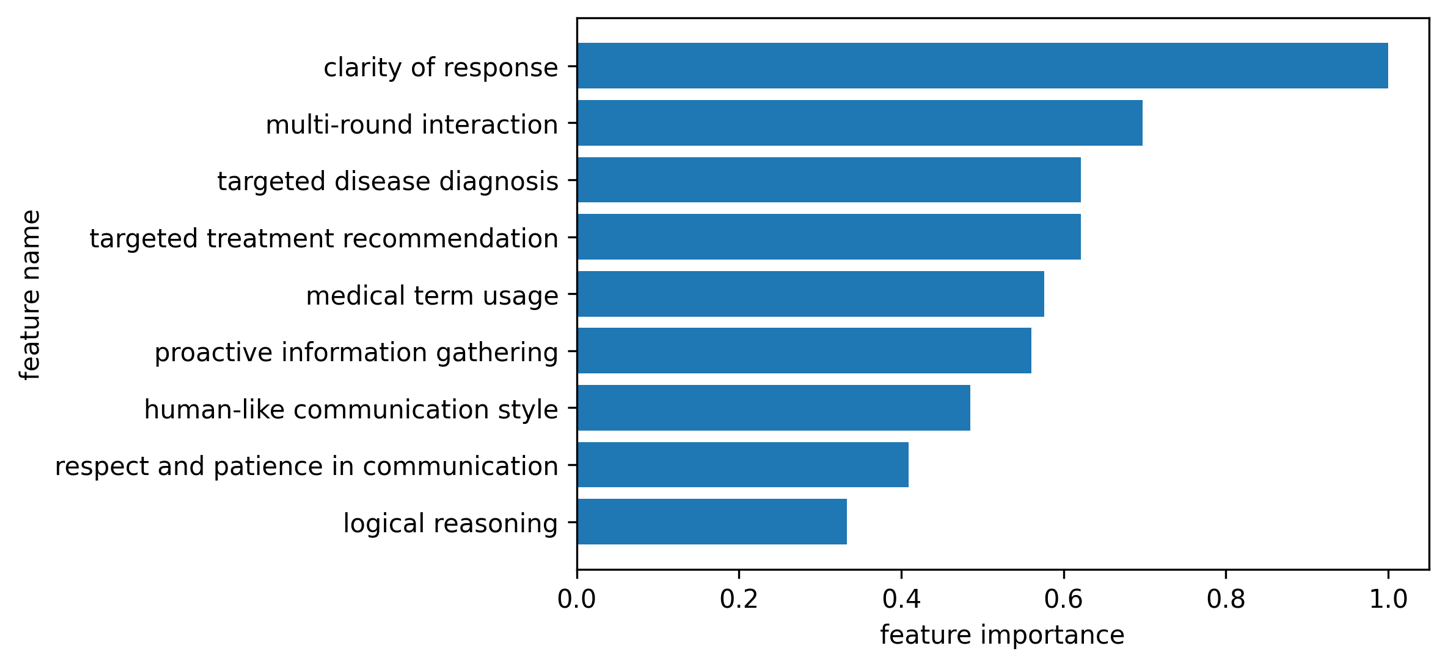}
         \caption{Trustworthiness}
     \end{subfigure}    
          \hfill
     \begin{subfigure}[b]{0.8\textwidth}
         \centering
         \includegraphics[width=\textwidth, height=5.5cm]{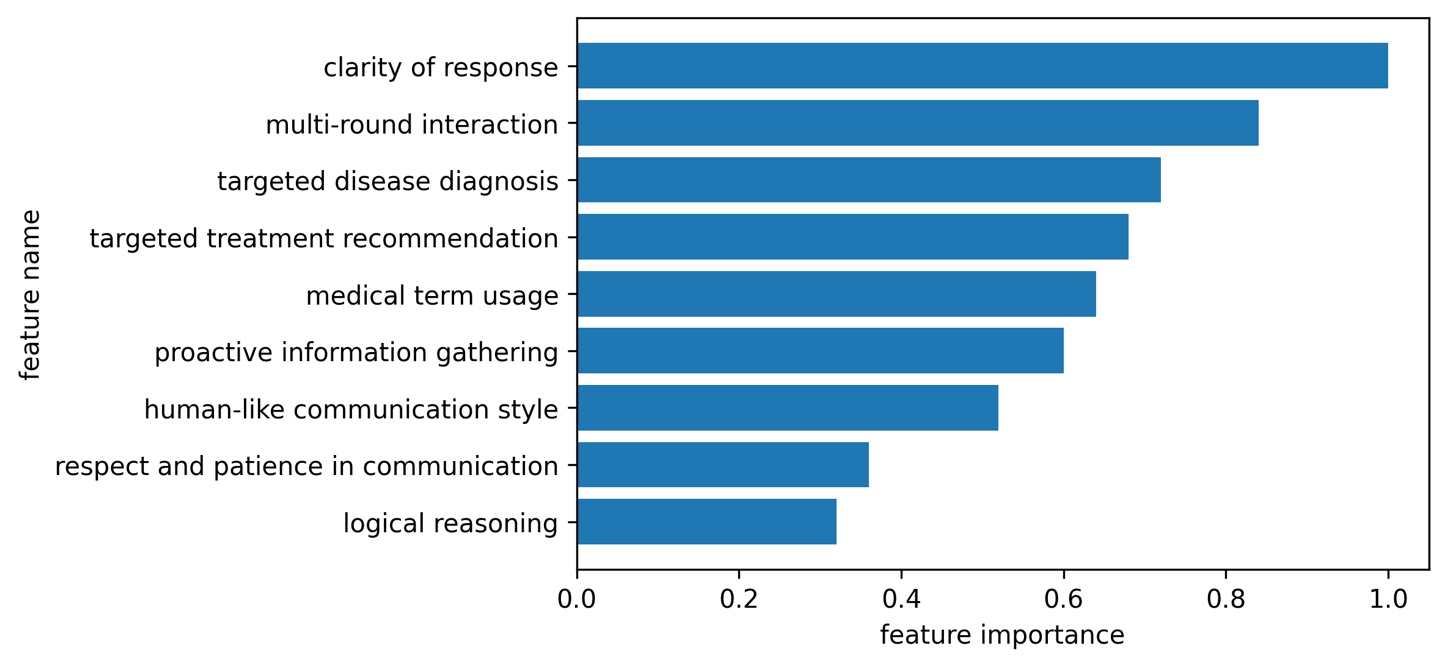}
         \caption{Overall Consumer Preference}
     \end{subfigure}    
        \caption{Feature Importance of Medical Consultation Record Characteristics on Consumer Preference} 
        \label{feature_importance_consumer_preference}
\end{figure}

We use the extracted textual characteristics to predict four metrics including professionalism, accuracy, consumer trustworthiness, and satisfaction, as well as overall medical expertise and consumer preference on all consultation records from three models. Medical expertise is calculated as the average of professionalism and accuracy, while consumer preference is the average of satisfaction and trustworthiness. We plot the learned feature importance for medical expertise in Figure~\ref{feature_importance_medical_expertise} and for consumer preference in Figure~\ref{feature_importance_consumer_preference}.

Interestingly, we find that the characteristics contributing to perceived medical expertise and consumer preference are very different.  For accuracy, the top five predictors are targeted disease diagnosis, medical term usage, proactive information gathering, multi-round interaction, and targeted treatment recommendation. The top predictors for perceived professionalism are human-like communication style, proactive information gathering, logical reasoning, multi-round interaction, and medical term usage. These aspects suggest that training and technology solutions should focus on enhancing these characteristics in medical professionals and AI systems to meet professional standards.

For consumer preference, the top predictors for consumer satisfaction are targeted treatment recommendations, respect and patience in communication, clarity of response, and medical term usage. For consumer trustworthiness, the top predictors are clarity of response, multi-round interaction, targeted disease diagnosis, targeted treatment recommendations, and medical term usage. This indicates that beyond technical medical competence, the manner in which information is communicated to patients is vital.
Managers in healthcare settings should, therefore, emphasize patient-centered communication skills in training programs to ensure clarity and respectfulness, tailoring interactions to effectively meet consumer needs. Our framework aligns the conversation pattern with that of a human doctor and models the soft skills and core principles into LLMs, which can improve consumer satisfaction and trustworthiness. These insights can guide the development of technology solutions that align with professional standards and consumer expectations, ultimately enhancing patient care and satisfaction.

\section{ Decision Support Showcase to Demonstrate the Practical Value to Assist Human Doctor}

In this section, we demonstrate the practical value of our model by illustrating how it can assist human doctors in the real world.

\subsection{Lab Experiments Design}
We showcase a decision support use case in which, during an online consultation, our LLM-doctor can be used to generate an initial response to patients' queries in a copilot manner. The human doctor can either accept this generated response or edit it before finalizing and sending it back to the patients. The copilot assistant approach enables human doctors to lead the entire process,  maintain control over the final response and conduct quality assurance, thereby mitigating potential errors or risky advice generated by the AI model. 

\begin{figure}[!ht]
     \centering
     \begin{subfigure}[b]{0.8\textwidth }
         \centering
         \includegraphics[width=\textwidth, height=8.5cm]{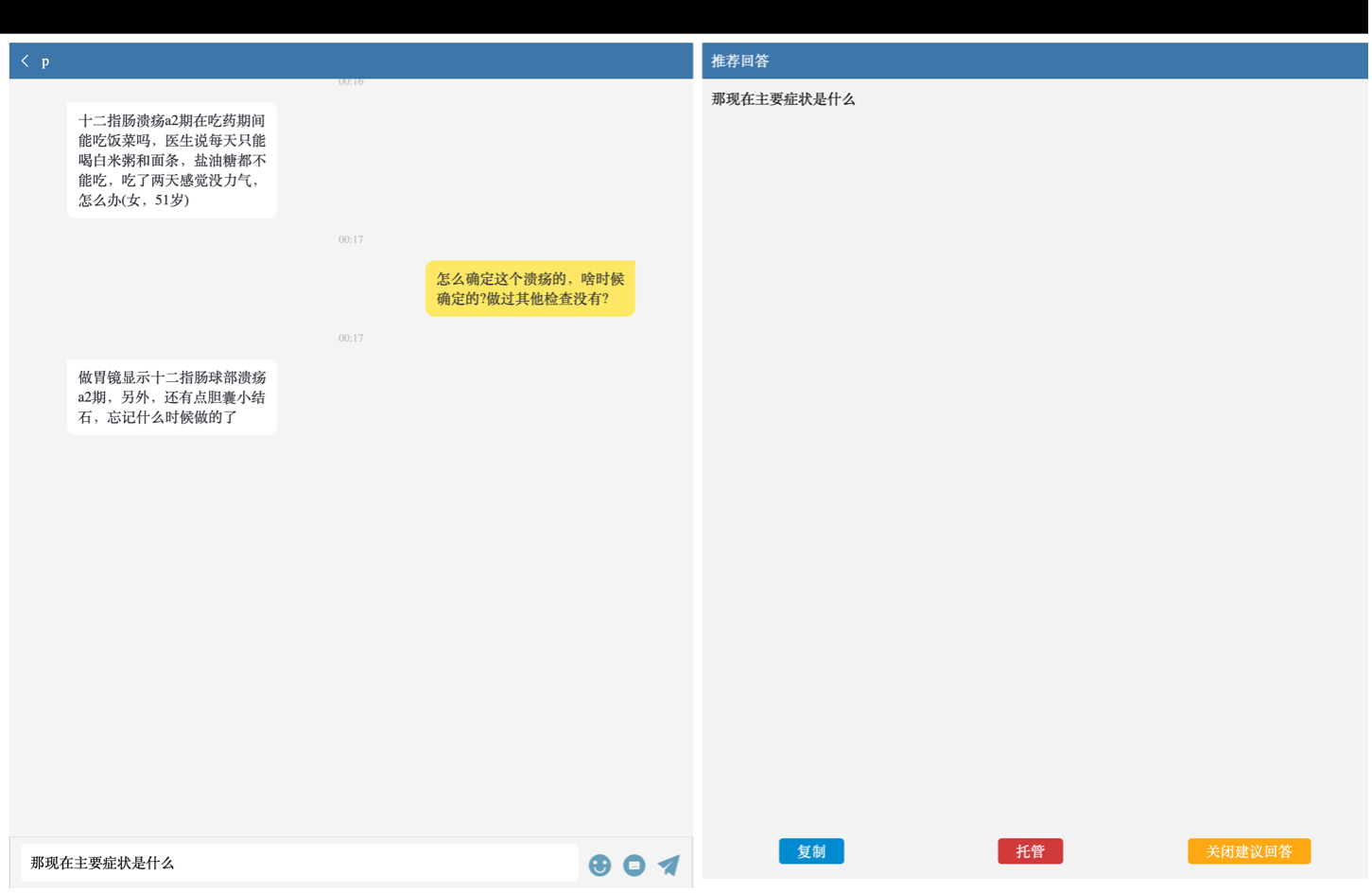}
         \caption{Interface with LLM-doctor Assistant}
     \end{subfigure}
     \hfill
     \begin{subfigure}[b]{0.8\textwidth}
         \centering
         \includegraphics[width=\textwidth, height=8.5cm]{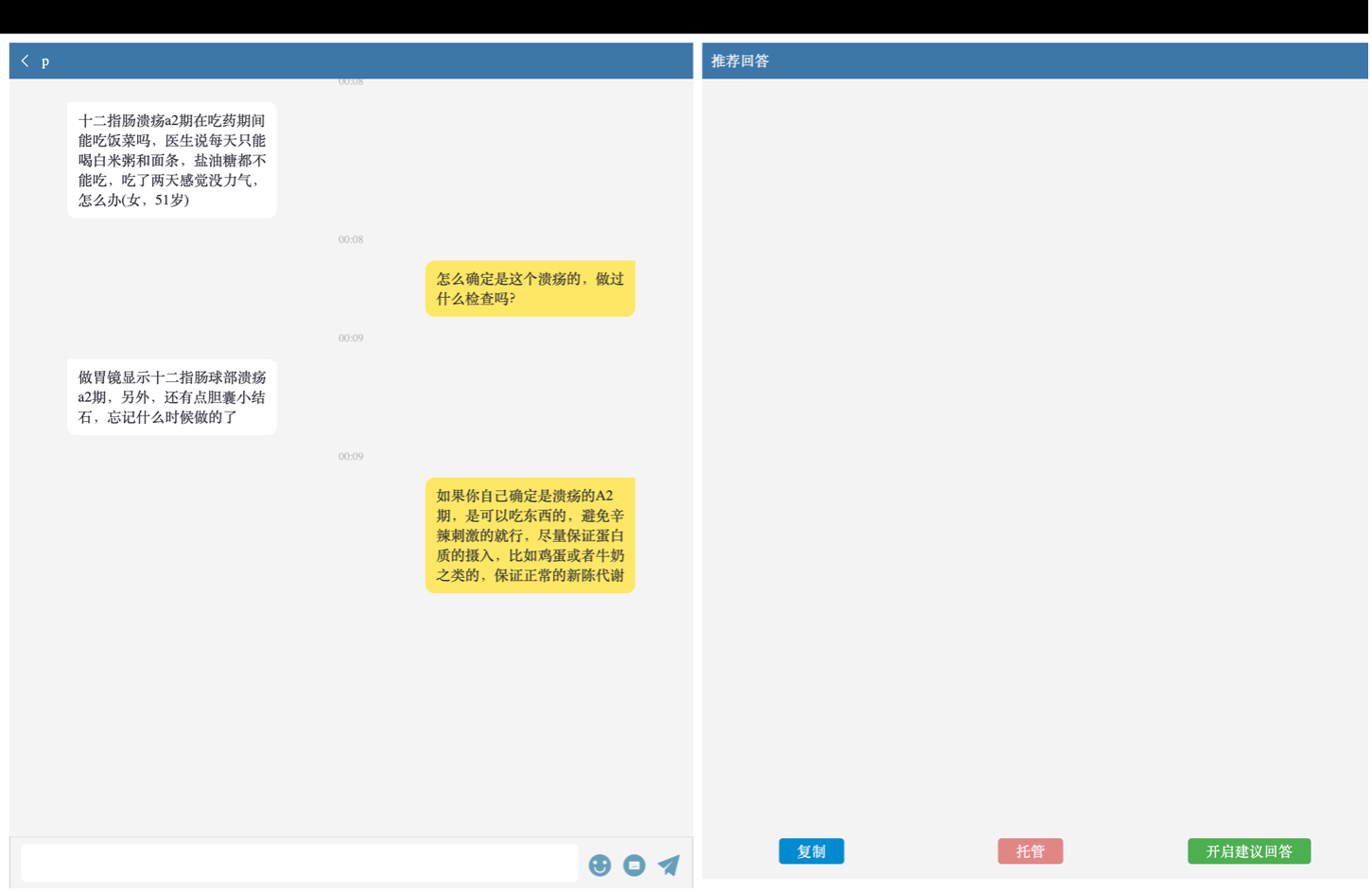}
         \caption{Interface without LLM-doctor Assistant}
     \end{subfigure}     
        \caption{Online Consultation Platform Interface} 
        \label{fig:interace}
\end{figure}

To achieve this goal, we develop an online consultation platform where our LLM-doctor or base model can be embedded in the backend to generate initial responses for the doctor in a copilot manner. For example, as illustrated in the screenshot in Figure~\ref{fig:interace}, when a patient submits a query or text, our model generates an initial response that appears in the chat box of the human doctor, who can then directly edit it in the chat box and send it to the patients. Such generation occurs in each round of interactions. Our platform supports a two-way connection, enabling both doctors and patients to log in simultaneously for consultations. It offers three modes: human doctor alone, human doctor + LLM-doctor, and human doctor + base model.

To demonstrate the practical value and effectiveness of such a system, we conduct a laboratory experiment, inviting five experienced doctors from prestigious hospitals in China. Each doctor participated in medical consultations with patients in all three modes, including human doctor alone, human doctor + LLM-doctor, and human doctor + base model. Allowing each doctor to experience all three modes helps them get a direct sense of how these models differ in their consultation process. To ensure a fair comparison, we recruit multiple research assistants and, according to the doctors' expertise, generate synthetic patient profiles from a random sample of historical records from Chunyu doctors using GPT-4. Each research assistant has a fixed patient profile and consults with the same human doctor three times under different modes. This synthetic patient profile avoids the risks of interacting with real patients and ensures a fair comparison of the three modes. We film the consultation process and save the recorded videos, which can help us analyze the entire consultation process. We also distribute a post-hoc survey to gather the doctors' evaluations of their  experience.

\subsection{Results }

Firstly, we compare the productivity of human doctors across the three modes as shown in Table~\ref{table:productivity}. We measure productivity by using the time spent by human doctors during consultations. Please note that we exclude patients' time spent, as our key focus is to observe the increase in human doctors' productivity. For each doctor, we show the time spent under three conditions and calculate the time saved by using the LLM-doctor and base model. With the aid of the LLM-doctor, human doctors, on average, saved 53.16\% of their time compared to operating without the LLM-doctor. The base model also saves human doctors' time. However, the magnitude is much smaller compared to the LLM-doctor; it saved only 19.31\% of the time compared to the doctors working independently. This suggests that our LLM-doctor can dramatically enhance the productivity of human doctors compared to the base model. Such a pattern is consistent for all human doctors in our experimental results.

\begin{figure}[!hbpt]
    \centering
    \includegraphics[scale = 0.35]{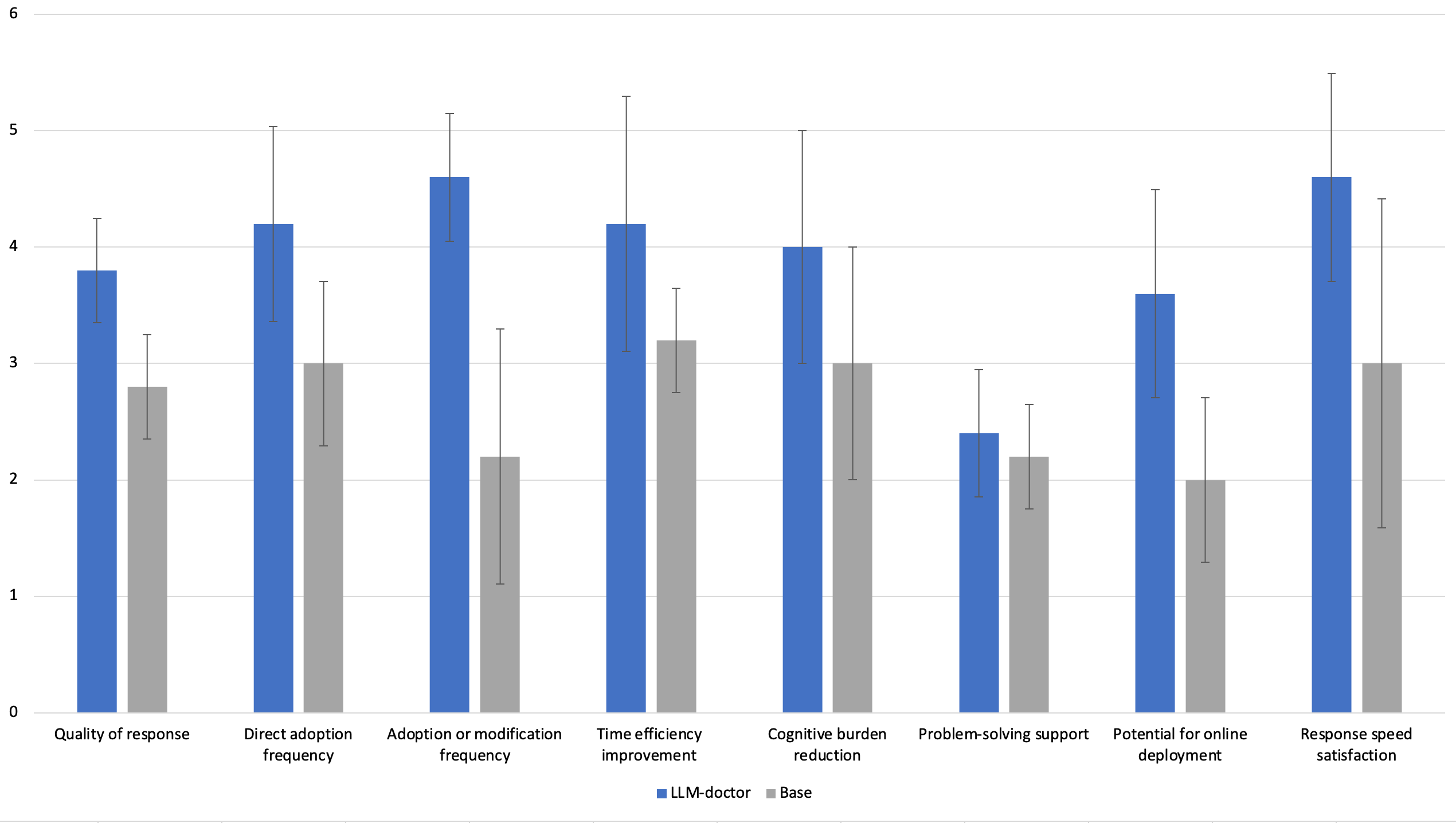}
    \caption{\centering  Human Doctors' Experience Evaluation Toward LLM-doctor and Base Model in Medical Consultation Lab Experiment}
    \label{survey}
\end{figure}

Furthermore, we collect post-hoc surveys from human doctors regarding the evaluation of their experiments and plot their responses in Figure~\ref{survey}. Specifically, we ask human doctors to rate the quality of response, frequency of direct adoption, frequency of adoption or modification, improvement in time efficiency, reduction in cognitive burden, support for problem-solving, potential for online deployment, and satisfaction with response speed on a 1-5 Likert scale, where 5 indicates extremely high and 1 indicates extremely low. We specifically pose the following questions to human doctors. Each doctor is asked to evaluate both the LLM-doctor and the base model using the same survey.

\begin{itemize}

    \item  \textit{Quality of response}: What do you think of the quality of AI-generated response suggestions?
    
    \item  \textit{Direct adoption frequency}: How frequently do you directly adopt AI-suggested answers in conversations with patients?
    
\item \textit{Adoption or modification frequency}: How frequently do you either directly adopt or modify (retaining at least 50\% of the original) AI-suggested answers in conversations with patients?

\item \textit{Time efficiency improvement}: Has the time you spend on patient consultations been reduced when using AI assistance?

\item \textit{Cognitive burden reduction}: Does AI assistance help reduce your cognitive workload?

\item \textit{Problem-solving support}: Can AI help you solve problems that you would find difficult to solve on your own?

\item \textit{Potential for online deployment}: How great do you think the potential is for large-scale deployment of this AI in online consultation platforms?

\item \textit{Response speed satisfaction}: Does the response speed of the AI meet your work needs?
\end{itemize}

Interestingly, the results show that human doctors generally perceive the response quality of the LLM-doctor to be higher, and they adopt its responses (either directly or by modifying, retaining at least 50\% of the original content) more frequently than those of the base model. Additionally, the doctors believe that the LLM-doctor can reduce their cognitive load during consultations much more than the base model. They also think the LLM-doctor can save them time spent during consultations more effectively than the base model. Furthermore, they perceive the LLM-doctor as having more practical value for large-scale deployment on online consultation platforms compared to the base model. These analyses show that the LLM-doctor's practical value in assisting human doctors is greater than that of the base model.

\begin{table}[tb]
\scriptsize
	\centering
\begin{tabular}{llll|llll}
\toprule
	& Human Doctor   &  Human Doctor  &  Human Doctor & Time Saved  & Time Saved  \\
 	& without AI  &  + LLM-doctor  &   + Base & by LLM-doctor & by Base  \\

 \midrule
Human Doctor 1 & 15.3 &	5.0 &	12.7 &  -67.32\% & 	-16.99\%\\
Human Doctor 2 & 11.5 &	3.5 &	8.3 & -69.57\% & 	-27.83\% \\
Human Doctor 3 & 10.7 &	2.9 &	10.0 & -72.90\% & 	-6.54\%\\
Human Doctor 4 & 9.4 &	6.8 &	8.2 & -27.66\%	& -12.77\%\\
Human Doctor 5 & 7.4 &	5.3 &	5 & -28.38\% & 	-32.43\%\\
\hline
Average & - & - & - & -53.16\%	& -19.31\%\\
\bottomrule
\end{tabular}
\caption{
Productivity Comparison: Time Spent (min) by Human Doctors under Different Experimental Conditions.}
\label{table:productivity}
\end{table}

Additionally, we also provide an open-ended question regarding their additional evaluation or comments toward our LLM-doctor and base model. For LLM-doctor, we receive the following open-ended evaluations and comments:
\begin{itemize}
    \item It is generally helpful and can often be used directly, although it occasionally provides responses that are too brief.
    \item It offers valuable insights that might be overlooked by doctors, thereby helping to decrease the risk of acute illnesses.
    \item The AI excels in proactively asking questions and effectively gathering information from patients, addressing the key points in its responses.
\end{itemize}

For the base model, we receive the following open-ended evaluations and comments:
\begin{itemize}

 \item Responses are too lengthy, verbose, and not immediately usable, necessitating manual input which increases the workload.
 \item Information lacks professionalism and personalization, and does not provide effective medical advice, making it less useful for both doctors and patients.
 \item The information provided is overly dense and generic and often includes many irrelevant details, similar to what is readily available online, adding little value for medical professionals and potentially increasing anxiety for patients with limited medical knowledge.
\end{itemize}

Overall, human doctors think the base model has salient drawbacks, including providing overly verbose and generic responses, a lack of personalization, and many irrelevant details which might increase anxiety for patients. These factors make it less useful for doctors and patients. In contrast, our LLM-doctor can proactively ask questions and address the key points in consultations and can be directly used by human doctors during consultations. All these factors demonstrate that our LLM-doctor's practical value is greater than that of the base model.

\section{Conclusion}

The significant progress in LLMs has profoundly impacted the conventional approach in design science research. This traditional approach involves identifying specific tasks, generating relevant datasets, and constructing models to connect inputs with outputs for these tasks. However, such approach is increasingly facing challenges due to the superior capabilities of LLMs.

In this study, we propose 
a novel framework to customize LLMs for general business contexts that aims to achieve three fundamental
objectives simultaneously: (1) aligning conversational patterns, (2) integrating in-depth domain knowledge,
and (3) embodying theory-driven soft skills and core principles. We design methodologies that combine
domain-specific theory with Supervised Fine Tuning (SFT) to achieve these objectives simultaneously. Our framework facilitates the customization of LLMs to serve
as general-purpose professional experts for business purposes, including demonstrating domain expertise and enhancing consumer satisfaction and trustworthiness.

Then, we adopt multiple experiments, including online experiment and lab experiment, to demonstrate the effectiveness and practical value of our framework in the context of medical consultations. Our evaluations include online experiments with actual patients and assessments by domain experts and real consumers show that customized LLM model substantially
outperforms untuned base model in medical expertise as well as consumer satisfaction and trustworthiness. Our framework significantly narrows the gap between untuned language models and human professionals, significantly moving closer to human-level performance.  In addition, we analyze the content of text-based consultation records and apply interpretable machine learning methods to identify the factors contributing to these performance improvements. We also demonstrate the practical applications of our model in a decision-support system intended to aid doctors during consultations in lab-based experiments. We implement an online platform for consultations where our model provides preliminary responses, supporting doctors in a supportive copilot fashion. The results indicate that with the support of the LLM-doctor, medical professionals can reduce their consultation time by an average of 53.16\%, and decrease their cognitive load, proving the model's high efficacy and adaptability for broad use in practice.

Our study carries fruitful managerial implications. Although we instantiate our framework in the context of medical consultation, our framework can be easily generalized to other business contexts. For example, it is possible to generalize it to customer support, legal assistance, sales and marketing, educational programs, and more. Our framework offers step-by-step principles,  guidance and valuable insights that future research can potentially use.

Furthermore, our study has practical implications for the healthcare industry. We demonstrate one possible way of integrating an LLM-doctor to assist human doctors in generating initial responses in a copilot manner. In fact, there are many other possibilities for integrating the customized LLM-doctor model into medical workflows, which can potentially benefit healthcare delivery by enhancing accessibility, scalability, and cost-effectiveness. It opens up new possibilities for 24/7 medical consultation services, especially in remote or underserved areas, and offers a solution to workforce shortages by handling routine consultations. This, in turn, allows medical professionals to concentrate on more complex cases, thereby improving the overall efficiency and effectiveness of healthcare systems.


\bibliography{ref}
\bibliographystyle{informs2014}

\end{document}